\documentclass[%
 reprint,
superscriptaddress,
groupedaddress,
 amsmath,amssymb,
 aps,
pra,
prb,
rmp,
prstab,
prstper,
]{revtex4-2}

\usepackage[utf8]{inputenc}
\usepackage{graphicx}
\usepackage{dcolumn}
\usepackage{bm}
\usepackage{array}
\usepackage{mathtools}
\usepackage{booktabs}
\usepackage{etoolbox,ragged2e}
\makeatletter
\patchcmd{\@makecaption}{\centering}{\RaggedRight}{}{}
\makeatother
\usepackage{algorithm}
\usepackage[noend]{algpseudocode}  
\usepackage{float}                  
\algrenewcommand\algorithmiccomment[1]{\hfill$\triangleright$~#1}

\usepackage[colorlinks=true,allcolors=blue,breaklinks=true]{hyperref}
\usepackage{booktabs} 
\usepackage{placeins}

\begin{document}

\preprint{APS/123-QED}

\title{Quasiparticle Variational Quantum Eigensolver}

\author{Saavanth Velury}
\email{velurys@ufl.edu}
\affiliation{Department of Physics, University of Florida, Gainesville, Florida 32611, USA}
\author{Yuxuan Wang}
\email{yuxuan.wang@ufl.edu}
\affiliation{Department of Physics, University of Florida, Gainesville, Florida 32611, USA}

\date{\today}

\begin{abstract}
We propose a momentum-space based variational quantum eigensolver (VQE) framework for simulating quasiparticle excitations in interacting quantum many-body systems on near-term quantum devices. Leveraging translational invariance and other symmetries of the Hamiltonian, we reconstruct the momentum-resolved quasiparticle excitation spectrum through targeted simulation of low-lying excited states using VQE. We construct a translationally symmetric variational ansatz designed to evolve a free-fermion particle-hole excited state with definite momentum $q$ to an excited state of the interacting system at the same momentum, employing a fermionic fast Fourier transform (FFFT) circuit coupled to a Hamiltonian Variational Ansatz (HVA) circuit. Even though the particle number is not explicitly conserved in the variational ansatz, the correct quasiparticle state is reached by energetic optimization. We benchmark the performance of the proposed VQE implementation on the XXZ Hamiltonian, which maps onto the Tomonaga-Luttinger liquid in the fermionic representation. Our numerical results show that VQE can capture the low-lying excitation spectrum of the bosonic quasiparticle/two-spinon dispersion of this model at various interaction strengths. We estimate the renormalized velocity of the quasiparticles by calculating the slope of the dispersion near zero momentum using the VQE-optimized energies at different system sizes, and demonstrate that it closely matches theoretical results obtained from Bethe ansatz. Finally, we highlight extensions of our proposed VQE implementation to simulate quasiparticles in other interacting quantum many-body systems.
\end{abstract}

\pacs{Valid PACS appear here}
\maketitle

\section{Introduction}

A long-standing objective for quantum computers has been to simulate quantum many-body systems~\cite{Feynman1982,Lloyd1996,Abrams1997}. However in the near term, quantum computers offer only modest qubit counts and short coherence times. Current quantum computers are incapable of error correction and lack fault tolerance, thus making the execution of many proposed quantum algorithms intractable. In order to take advantage of the quantum computers that exist in this current noisy intermediate scale quantum (NISQ) era~\cite{Preskill2018}, variational quantum algorithms (VQAs) have become the prevailing strategy for quantum simulation~\cite{Cerezo2021,Bharti2022}, such as the variational quantum eigensolver (VQE)~\cite{Peruzzo2014,Wecker2015a,Bauer2016,Tilly2022}. Working within the constraints of NISQ devices, these algorithms utilize an optimization-based technique that is executed on a classical computer alongside an objective function prepared on a quantum computer representing the quantum system of interest, thereby realizing a hybrid quantum-classical workflow. By leveraging the resources of both classical and quantum computers, VQAs serve as an interim, hybrid paradigm for simulating physical systems, offering practical traction until fault-tolerant quantum devices are realized to execute more powerful quantum algorithms at scale.

In condensed matter physics, VQE has been adopted in simulating the ground states of strongly correlated electronic systems such as the Fermi-Hubbard model~\cite{Babbush2014,Hastings2015,Wecker2015,Wecker2015a,McClean2016,Babbush2018,Jiang2018,Kivlichan2018,Reiner2019,Cade2020,AnselmeMartin2022,Gyawali2022,Stanisic2022,Jafarizadeh2024,Bespalova2025,Alvertis2025}, spin liquids~\cite{Bespalova2021,Feulner2022,Jahin2022,Li2023,Kattemolle2022} and topological phases~\cite{Sun2023,Okada2023,Roy2024,Ciaramelletti2025,Shen2025}. Two components of the VQE algorithm that can have a substantial impact on its performance include (i) initial state preparation and (ii) variational ansatz construction. Refinements to initial-state preparation primarily involve preparing quantum states that have significant overlap with target eigenstates of the Hamiltonian. For example, initializing a physically motivated reference state (e.g., Hartree-Fock / mean-field state) can restrict the search space for optimization and thus accelerate convergence~\cite{Wecker2015,Wecker2015a,Tubman2018,Gard2020,Selvarajan2022,Fomichev2024,Berry2025,GreeneDiniz2025}. Furthermore, designing a variational ansatz that preserve the symmetries of the Hamiltonian can improve performance by constraining the variational manifold to the relevant symmetry sector with a fixed set of quantum numbers, thus mitigating leakage into other subspaces during optimization and potentially allowing for shallower circuits~\cite{Barron2021,Bertels2022,Mihalikova2025,Shirali2025}.

Although most VQE studies have focused primarily on ground state simulation, there has also been growing interest in simulating excited states~\cite{Nakanishi2019,Higgott2019,Ryabinkin2019,Jones2019,Carobene2023,Gocho2023,Lyu2023,Yao2025}. In this work, we focus on a particular class of excited states that are long-lived, particle-like collective modes known as quasiparticles~\cite{Bohm1951,Pines1952,Abrikosov1975,Coleman2015}. In interacting systems, quasiparticles emerge from collective motion and are characterized by a dispersion relation and well-defined quantum numbers (e.g., spin, momentum), with parameters renormalized by interactions (such as effective mass or velocity, and coupling to charge when present). Quasiparticles are a key notion of condensed matter physics, which provide a unifying framework for understanding a wide variety of emergent phenomena, ranging from phonons and magnons to Fermi-liquid excitations and anyons in topological phases. Their ubiquity across condensed matter systems makes them a natural target for quantum simulation. Recently, many proposals have been put forth to simulate quasiparticles in strongly correlated systems on near-term quantum devices. These include quantum subspace expansion techniques~\cite{Yoshioka2022,Ohgoe2024,Umeano2025,Ohgoe2025} and Green's function methods~\cite{Endo2020,Kosugi2020,Libbi2022,Gyawali2022,Bishop2025,VilchezEstevez2025,Piccinelli2025,Umeano2025}. There have also been more specialized and unique approaches to quasiparticle simulation using VQE. These include Ref.~\cite{Seki2020}, which proposed projecting the VQE-optimized circuit state into a target symmetry sector in classical post-processing, Ref.~\cite{Chen2025} which proposed a tangent space excitation ansatz to capture low-lying excitations, Ref.~\cite{Jaiswal2025} which (for non-interacting quasiparticles) proposed the preparation of Wannier states with VQE and exploiting translation symmetry to reconstruct the momentum-resolved quasiparticle energy dispersion, Ref.~\cite{Falide2025} which performed VQE using a single-spinon ansatz to reconstruct the spinon energy dispersion for spin-1/2 systems, and Ref.~\cite{Sumeet2025} which combines numerical linked-cluster expansion techniques with VQE to compute the quasiparticle energy dispersion.

In this work, we present a momentum-resolved, symmetry-preserving VQE framework  that encodes the quantum numbers of the quasiparticles in the ansatz choice. We alternate between momentum and real space: the initial state is constructed in momentum space to ensure the correct quantum numbers and then transformed to real space for the variational circuits, where the interactions are local.  As variational methods are designed to solve the eigenstates with the lowest energy, they cannot extract much information about a continuous spectrum. However, for long-lived quasiparticles with sharp spectral functions, VQE is particularly effective. To that end, we adopt a targeted strategy: for each subspace labeled by momentum $q$, we initialize the matching free-fermion eigenstate and run VQE to obtain the corresponding excited energy. Performing VQE for different small values of $q$ then yields the momentum space energy dispersion. We find that while it is possible to ensure the variational circuits preserve both momentum and fermion number, it is much more efficient to only preserve momentum. This is because introducing additional fermionic particles/holes costs extra energy. Therefore, by solely optimizing the energy, the fermion number of the initial state is preserved. 

To demonstrate the effectiveness of our framework, we use the XXZ model as an example, a spin chain that is closely related to the Tomonaga-Luttinger liquid~\cite{Luttinger1963,Mattis1965,Luther1975,Giamarchi2003} hosting bosonic quasiparticle excitations with relativistic energy dispersion. We perform classical statevector simulations and demonstrate that our momentum space based VQE framework can recover the quasiparticle energy dispersion $E(q)$. Then, we estimate the velocity of the excitations by computing the slope of the dispersion near zero momentum using the VQE-optimized energies at various system sizes and show that they are in good agreement with established results from Bethe ansatz.

This paper is organized as follows. In Sec.~\ref{sec:overview} we provide a review on the background and methods used in this work, with Sec.~\ref{subsec:XXZ} going over the properties of the XXZ spin chain, Sec.~\ref{subsec:HVAVQE} providing a high-level overview of the ansatz construction and VQE framework, and Sec.~\ref{subsec:ansatz} going over the quantum circuit implementation. Then in Sec.~\ref{sec:results}, we discuss the performance of VQE for the XXZ model, providing a detailed presentation of the numerical results including plots of the quasiparticle energy dispersion, relative energy and fidelity errors, and computed values for the effective velocity of the quasiparticle excitations. Finally, we conclude in Sec.~\ref{sec:conclusion}, and highlight extensions of this framework to other interacting quantum many-body systems.

\section{Background and methods}
\label{sec:overview}

We focus on developing and benchmarking a symmetry-preserving, momentum-resolved VQE framework that generates quasiparticle excitations. The spin-1/2 XXZ chain is an ideal testbed for this purpose: it is Bethe-ansatz integrable~\cite{Bethe1931}, providing analytical results that serve as stringent benchmarks to compare against our VQE results. Furthermore, its low-energy sector maps to a Tomonaga-Luttinger liquid with linearly dipsersing bosonic quasiparticles characterized by a renormalized velocity $v$. Although quantum circuits for the exact realization of Bethe ansatz states of the XXZ model have been proposed~\cite{VanDyke2021,Li2022,Sopena2022,Raveh2024,Ruiz2024,Ruiz2025}, some of these methods are probabilistic in their implementation and require non-unitary steps such as post-selection, which makes them computationally expensive at scale and may not be suitable for implementation on NISQ devices. Therefore, assessing how well our VQE framework performs in reproducing the low-lying quasiparticle energy dispersion is thus both practically relevant for implementation on near-term quantum computers and informative for extensions to non-integrable models and fermionic strongly correlated systems.

\subsection{Review of XXZ Model}
\label{subsec:XXZ}

The XXZ Hamiltonian with periodic boundary conditions is given by (for definiteness, assuming that the number of sites $N$ is even and $n\in\left\{-\frac{N}{2}+1,\ldots,\frac{N}{2}\right\}$)
\begin{equation}\begin{split}\label{eq:XXZHamI}
\mathcal{H}_{\text{XXZ}}&=\sum_{n}\left(\sigma_{n}^{x}\sigma_{n+1}^{x}+\sigma_{n}^{y}\sigma_{n+1}^{y}+\Delta\sigma_{n}^{z}\sigma_{n+1}^{z}\right)\\
&\equiv\mathcal{H}_{\text{XX}}+\Delta\mathcal{H}'.
\end{split}\end{equation}
In the above equation, we have decomposed the XXZ Hamiltonian into two terms: the XX Hamiltonian ($\mathcal{H}_{\text{XX}}$) and an Ising term ($\mathcal{H}'$). The parameter $\Delta$ controls the anisotropy between these two terms. The reason for this decomposition is apparent when mapping it to the fermionic basis via the Jordan-Wigner transformation (see Appendix~\ref{sec:JWTransform} for a review),
\begin{equation}\begin{split}\label{eq:XXZHamII}
&\mathcal{H}_{\text{XX}}=2\sum_{n}\left(c_{n}^{\dagger}c_{n+1}+\text{h.c.}\right)+\mathcal{H}_{\text{boundary}}\\
&\mathcal{H}'=\sum_{n}\left(2c_{n}^{\dagger}c_{n}-1\right)\left(2c_{n+1}^{\dagger}c_{n+1}-1\right).
\end{split}\end{equation} 
The boundary term is given as
\begin{equation}\begin{split}\label{eq:XXZboundary}
\mathcal{H}_{\text{boundary}}=-2\mathcal{P}\left(c_{\frac{N}{2}}^{\dagger}c_{-\frac{N}{2}+1}+\text{h.c.}\right).
\end{split}\end{equation}
In the above, we have introduced the fermion parity operator $\mathcal{P}$ (not to be confused with spatial inversion),
\begin{equation}\begin{split}
\mathcal{P}=\exp\left\{i\pi\mathcal{N}_{F}\right\}=(-1)^{\mathcal{N}_{F}}=\prod\limits_{n=-\frac{N}{2}+1}^{\frac{N}{2}}\sigma_{n}^{z}
\end{split}\end{equation}
which commutes with $\mathcal{H}_{\text{XXZ}}$ and $\mathcal{H}_{\text{XX}}$ ($[\mathcal{P},\mathcal{H}_{\text{XXZ}}]=[\mathcal{P},\mathcal{H}_{\text{XX}}]=0$). Effectively, for an even $N$, the fermions satisfy periodic boundary conditions at even filling and antiperiodic ones at odd filling. Intuitively, this prevents any ambiguity in filling the lowest energy states in the Fermi sea, which is unphysical from the perspective of the spin model. Crucially, the fermion parity determines the boundary conditions of the Hamiltonian in the fermionic basis, although the effects of this term on the spectrum can be safely neglected in the thermodynamic limit~\cite{Lieb1961}. 

For $\Delta\neq 1$, the XXZ model is symmetric under a $U(1)$ spin rotation around the $z$ axis. Under the fermionic representation, the $U(1)$ spin rotation symmetry is mapped to the $U(1)$ symmetry of the fermions. This leads to a mapping between the $U(1)$ charges in both representations: namely the total fermion number $\mathcal{N}_{F}$ and the $z$-component of the total spin, $S_z$, are related by 
\begin{equation}\begin{split}
\mathcal{N}_{F}=\sum\limits_{n=-\frac{N}{2}+1}^{\frac{N}{2}}c_{n}^{\dagger}c_{n}\equiv S_{z}+\frac{N}{2}.
\end{split}\end{equation}
The system also preserves particle-hole symmetry
\begin{equation}\begin{split}\label{eq:chargeconjugation}
\mathcal{C}:\hspace{0.1cm}c_{j}\to(-1)^{j}c_{j}^{\dagger}
\end{split}\end{equation}
corresponding to $S_z\to -S_z$, as well as the translation symmetry
\begin{equation}\begin{split}
\mathcal{T}:\hspace{0.1cm}c_{j}\to c_{j+1}\hspace{0.1cm}\left(c_{j}^{\dagger}\to c_{j+1}^{\dagger}\right).
\end{split}\end{equation}
Additionally, the system also exhibits spatial inversion symmetry $\mathcal{I}$, which maps the lattice site $n$ to $N+1-n$. The eigenstates of the XXZ Hamiltonian can be labeled by irreducible representations of these symmetries, including momentum, parity, and fermion number (or equivalently, the total $z$-component of spin).

In the fermionic basis, the XXZ Hamiltonian describes a lattice model of interacting spinless fermions. Taking the fermion parity to be $\mathcal{P}=-1$ to enforce periodic boundary conditions, the XXZ Hamiltonian can be expressed as follows (up to an overall constant),
\begin{equation}\begin{split}\label{eq:fermionicHam}
&\mathcal{H}_{\text{XXZ}}=2\sum_{n}\left(c_{n}^{\dagger}c_{n+1}+\text{h.c.}\right)-\mu\sum_{n}c_{n}^{\dagger}c_{n}\\
&+U\sum_{n}c_{n}^{\dagger}c_{n}c_{n+1}^{\dagger}c_{n+1}
\end{split}\end{equation}
where $\mu$ and $U$ are the parameters for the chemical potential and interaction strength respectively, with $\mu=U=4\Delta$. Note that the $\mu$ term is needed to ensure particle-hole symmetry, which would have been broken by the Hartree shift from the $U$ term. The value of $\mu$ can be further tuned away from the particle-hole symmetric point by adding an external field $h\sum_n \sigma^z_{n}$, which can be used to probe the $S_z \neq 0$ sector.

\begin{figure}
\includegraphics[width=0.5\textwidth]{"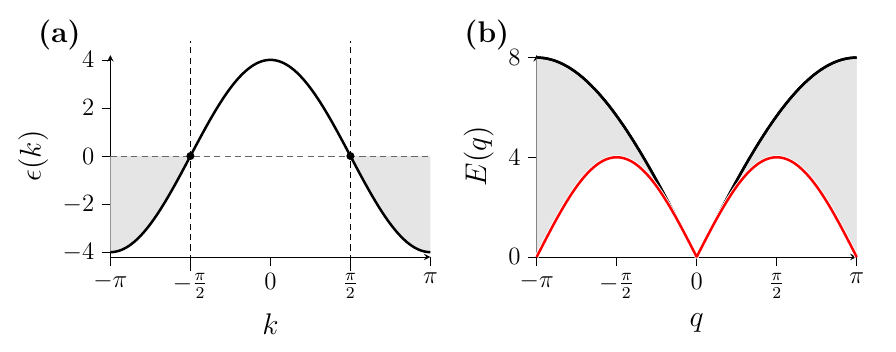"}
\caption{(a) Energy dispersion of the XX model, $\epsilon(k)=4\cos\,k$, with the Fermi points at $k_{F}=\pm\frac{\pi}{2}$. The shaded area indicates the filled negative energy states of the ground state (half-filled sector). (b) Spectrum of particle-hole excitations in the XX model in the half-filled sector. The bounding curves are highlighted in black and red, and the shaded area is the continuum representing the different excitation energies possible, with $q=k_{p}-k_{h}$.}
\label{fig:XXSpectra}
\end{figure}

When $\Delta=0$, Eq.~\eqref{eq:fermionicHam} reduces to $\mathcal{H}_{\text{XX}}$ which, as shown in Eq.~\eqref{eq:XXZHamII}, is a simple tight-binding model for a spinless electron with nearest-neighbor hoppings. The energy dispersion is given by $\epsilon(k)=4\cos\,k$ and is illustrated in Fig.~\ref{fig:XXSpectra}(a). The Fermi points are located at $k_{F}=\pm\frac{\pi}{2}$, and the ground state corresponds to the negative energy states being completely filled (i.e., the Fermi energy is $E_{F}=0$), as shown by the shaded area in gray in Fig.~\ref{fig:XXSpectra}(a). Performing the Fourier transform on $\mathcal{H}_{\text{XX}}$, the ground state can be expressed as
\begin{equation}\begin{split}\label{eq:XXGS}
|\Psi_{\text{GS}}\rangle=\prod\limits_{|k|>\frac{\pi}{2}}\tilde{c}_{k}^{\dagger}|\tilde{0}\rangle,
\end{split}\end{equation}
where $\tilde{c}_{k}^{\dagger}$ is the fermionic creation operator in momentum space. An excitation is formed by annihilating an electron at momentum $|k_{h}|>\frac{\pi}{2}$ and creating an electron above the Fermi sea at momentum $|k_{p}|<\frac{\pi}{2}$, known as a particle-hole excitation. Denoting $k_{h}\equiv k$ and $k_{p}-k_{h}\equiv q$ as the transfer momentum, the spectrum of particle-hole excitations can be determined by computing
\begin{equation}\begin{split}
E(q)=\epsilon(k+q)-\epsilon(k).
\end{split}\end{equation}
For a given $q$, a range of excitation energies is possible, depending on the value of $k$. This is illustrated in Fig.~\ref{fig:XXSpectra}(b), with the shaded gray area representing the particle-hole continuum, and the spectrum is completely gapless. The upper and lower bounding curves, highlighted in black and red respectively, are given by $E_{\text{max}}(q)=8\left|\sin\,\frac{q}{2}\right|$ and $E_{\text{min}}(q)=4\left|\sin\,q\right|$. At $q\to 0$, the two curves merge and the spectral function becomes sharp, corresponding to long-lived linearly-dispersing bosonic quasiparticles.

For $\Delta\neq 0$, fermionic interactions are induced and the low-energy sector of the XXZ chain maps onto the Tomonaga-Luttinger liquid. For $\Delta<1$, the system remains gapless, while for $|\Delta|>1$ the system becomes Ising-like and gapped. In the gapless phase, the low-lying excitations are long-lived bosonic quasiparticles in a Tomonaga-Luttinger liquid, which represent density fluctuations (charge in the context of interacting fermions, spin in the context of spin-chains). The excitation spectrum is qualitatively similar to the particle-hole excitation spectrum shown in Fig.~\ref{fig:XXSpectra}(b): at small $q$, the collective modes are linearly dispersing, characterized by a renormalized velocity $v$. Using the Bethe ansatz, the direct relationship between the XXZ Hamiltonian and the Luttinger liquid demonstrates that quantities such as the renormalized velocity $v$ can be applied to the XXZ Hamiltonian to characterize its low-energy excitations~\cite{Sirker2012},
\begin{equation}\begin{split}\label{eq:theoreticalvelocity}
v=2\pi\frac{\sqrt{1-\Delta^{2}}}{\cos^{-1}\,\Delta}.
\end{split}\end{equation}

Although we focus on finite-size systems for this work, Eq.~\eqref{eq:theoreticalvelocity} will serve as a useful benchmark to compare against the low-lying excitation spectrum of states obtained from VQE. 

\subsection{Ansatz Design \& VQE}\label{subsec:HVAVQE}

\begin{figure}
\includegraphics[width=0.49\textwidth]{"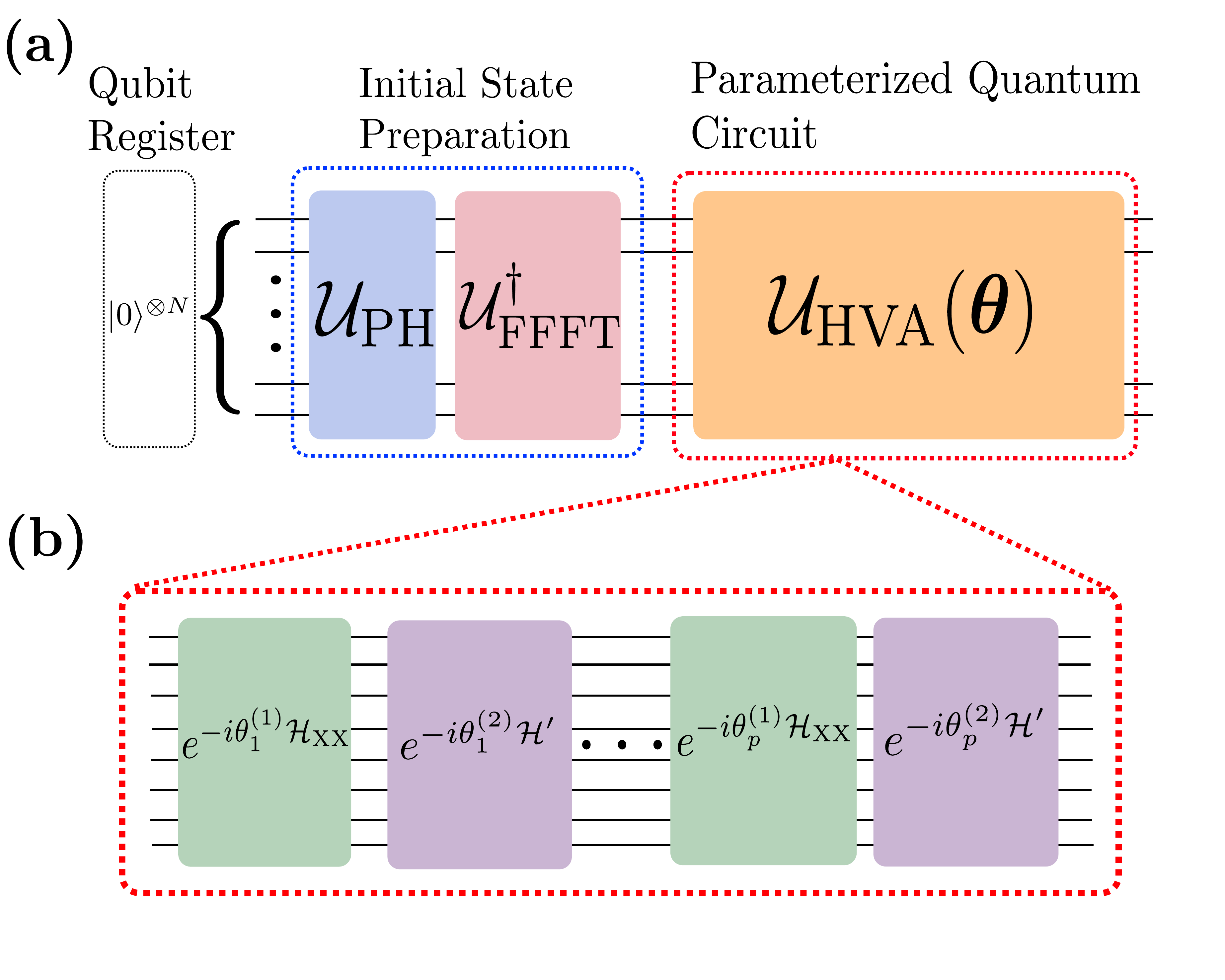"}
\caption{(a) High-level illustration of the variational ansatz circuit structure, comprised of the free-fermion state preparation circuit $\mathcal{U}_{\text{PH}}$, the inverse fermionic fast Fourier transform (FFFT) circuit $\mathcal{U}_{\text{FFFT}}^{\dagger}$, and the Hamiltonian Variational Ansatz (HVA) circuit $U_{\text{HVA}}(\boldsymbol{\theta})$. (b) Displays the Trotterized structure of the HVA circuit for $p$ layers and parameter vector $\boldsymbol{\theta}=\begin{pmatrix}\theta_{1}^{(1)}, & \theta_{1}^{(2)}, & \cdots & \theta_{p}^{(1)}, & \theta_{p}^{(2)}\end{pmatrix}$.}
\label{fig:VQPAoverview}
\end{figure}

Our ansatz utilizes eigenstates of $\mathcal{H}_{\text{XX}}$ as input states. This is then fed into a parameterized quantum circuit that variationally maps an eigenstate of $\mathcal{H}_{\text{XX}}$ to an eigenstate of $\mathcal{H}$ while preserving the symmetries shared by both $\mathcal{H}_{\text{XX}}$ and $\mathcal{H}$. The overall circuit structure is shown in Fig.~\ref{fig:VQPAoverview}. First, the qubit register is initialized in the vacuum state of the Fock space, which is represented by all qubits in the $|0\rangle$ state. The overall circuit structure can then be distilled into three components that are each denoted by unitaries: (i) a momentum-space free-fermion state preparation circuit $\mathcal{U}_{\text{PH}}$, (ii) an inverse fast fermionic Fourier transform circuit $\mathcal{U}_{\text{FFFT}}$ converting the basis to real space, and (iii) the Hamiltonian Variational Ansatz (HVA) circuit $\mathcal{U}_{\text{HVA}}(\boldsymbol{\theta})$. 
\begin{enumerate}
\item\textit{Momentum-Space Free-Fermion State Preparation $\mathcal{U}_{\text{PH}}$:} This circuit is designed to initialize a free-fermion state of $\mathcal{H}_{\text{XX}}$ in momentum space. Due to the symmetries shared by $\mathcal{H}_{\text{XX}}$ and $\mathcal{H}_{\text{XXZ}}$, the initial state is labeled by $q$ (total momentum) and $\mathcal{N}_{F}$ (fermion number). 
\item\textit{Inverse FFFT $\mathcal{U}_{\text{FFFT}}^{\dagger}$:} This circuit maps the prepared initial state from $k$-space to real space. The reason for doing this is that the interactions are local, and the variation of the initial state can be performed more efficiently in real space. We discuss this in further detail in Sec.~\ref{subsec:ansatz}.
\item\textit{Hamiltonian Variational Ansatz $\mathcal{U}_{\text{HVA}}(\boldsymbol{\theta})$:} Heuristically, this parameterized quantum circuit can be  expressed as
\begin{equation}\begin{split}\label{eq:HVA}
\mathcal{U}_{\text{HVA}}(\boldsymbol{\theta})\sim\prod\limits_{\ell=1}^{p}e^{-i\theta_{1}^{(\ell)}\mathcal{H}_{\text{XX}}}e^{-i\theta_{2}^{(\ell)}\mathcal{H}'}
\end{split}\end{equation}
where $p$ is the number of layers and $\boldsymbol{\theta}=\begin{pmatrix}\theta_{1}^{(1)}, & \theta_{2}^{(1)}, & \cdots & \theta_{1}^{(p)}, & \theta_{2}^{(p)}\end{pmatrix}$ is the parameter vector. The structure of Eq.~\eqref{eq:HVA} resembles a Trotter decomposition typically employed in time evolution, with the evolution time intervals replaced by the variational parameters contained in $\boldsymbol{\theta}$. The goal of the HVA is to map an input state, in this case the free-fermion eigenstate of $\mathcal{H}_{\text{XX}}$, to an eigenstate of $\mathcal{H}_{\text{XXZ}}$, through an alternating sequence of parameterized unitaries constructed from $\mathcal{H}_{\text{XX}}$ and $\mathcal{H}'$. Since $\mathcal{H}'$ represents a two-body operator in the fermionic basis, each application of $e^{-i\theta_{1}^{(\ell)}\mathcal{H}_{\text{XX}}}$ followed by $e^{-i\theta_{2}^{(\ell)}\mathcal{H}'}$ to the initial state adiabatically introduces interactions, eventually resulting in convergence toward an eigenstate of $\mathcal{H}_{\text{XXZ}}$. The depth $p$ determines the number of variational parameters to be optimized classically, analogous to the timescale over which the adiabatic evolution takes place; increasing the number of layers in the Trotter decomposition can result in a more accurate estimate. It should be emphasized that Eq.~\eqref{eq:HVA} is heuristic and is intended to demonstrate the physical motivation behind the structure of the HVA. At the gate level, directly exponentiating the many-body Hamiltonian 
$\mathcal{H}_{\text{XX}}$ is impractical. Instead, we Trotterize 
$e^{-i\theta_{1}^{(\ell)}\mathcal{H}_{\text{XX}}}$ into a product of 
two-qubit operators. However, this procedure tends to break either the 
$U(1)$ or translational symmetries. We choose to enforce translation symmetry and describe the circuit implementation of Eq.~\eqref{eq:HVA} in further detail in Sec.~\ref{subsec:hva}.

\end{enumerate}
Putting all three components together, the ansatz $|\Psi(\boldsymbol{\theta})\rangle$ in its complete form can be expressed as
\begin{equation}\begin{split}\label{eq:finalansatz}
|\Psi(\boldsymbol{\theta})\rangle=\mathcal{U}_{\text{HVA}}(\boldsymbol{\theta})\mathcal{U}_{\text{FFFT}}^{\dagger}\mathcal{U}_{\text{PH}}|0\rangle
\end{split}\end{equation}
where we have used $|0\rangle$ as a shorthand for the initialized qubit register $|0\rangle^{\otimes N}$.

Finally, we implement the standard hybrid quantum-classical framework for VQE, tailored to our symmetry-resolved setting. For a fixed momentum $q$ and fermion parity $\mathcal{P}$ with the ansatz $|\Psi(\boldsymbol{\theta})\rangle$ given by Eq.~\eqref{eq:finalansatz}, we minimize the variational energy $E(\boldsymbol{\theta})=\langle\Psi(\boldsymbol{\theta})|\mathcal{H}_{\text{XXZ}}|\Psi(\boldsymbol{\theta})\rangle$ which serves as the objective function for optimization. The steps of the VQE pipeline are as follows:
\begin{enumerate}
\item $|\Psi(\boldsymbol{\theta})\rangle$ is constructed on a quantum computer. 
\item The objective function $E(\boldsymbol{\theta})$ is then evaluated by measuring the expectation value of $\mathcal{H}_{\text{XXZ}}$ with respect to $|\Psi(\boldsymbol{\theta})\rangle$ on a quantum computer. 
\item An optimizer is run on a classical computer which updates the parameters $\boldsymbol{\theta}$ based on $E(\boldsymbol{\theta})$'s returned from the quantum computer, in order to lower $E(\boldsymbol{\theta})$. 
\end{enumerate}
These steps are repeated iteratively until the optimizer converges and yields a set of optimal parameters $\boldsymbol{\theta}^*$ and the corresponding VQE-optimized state $|\Psi(\boldsymbol{\theta}^*)\rangle$. In our framework, this yields the lowest-energy eigenstate within the targeted $(q,\mathcal{P})$ sector. The specific details of this implementation are discussed in Sec.~\ref{sec:results}.

\subsection{Implementing the Initial State}\label{subsec:ansatz}

\begin{figure*}
\includegraphics[width=\textwidth]{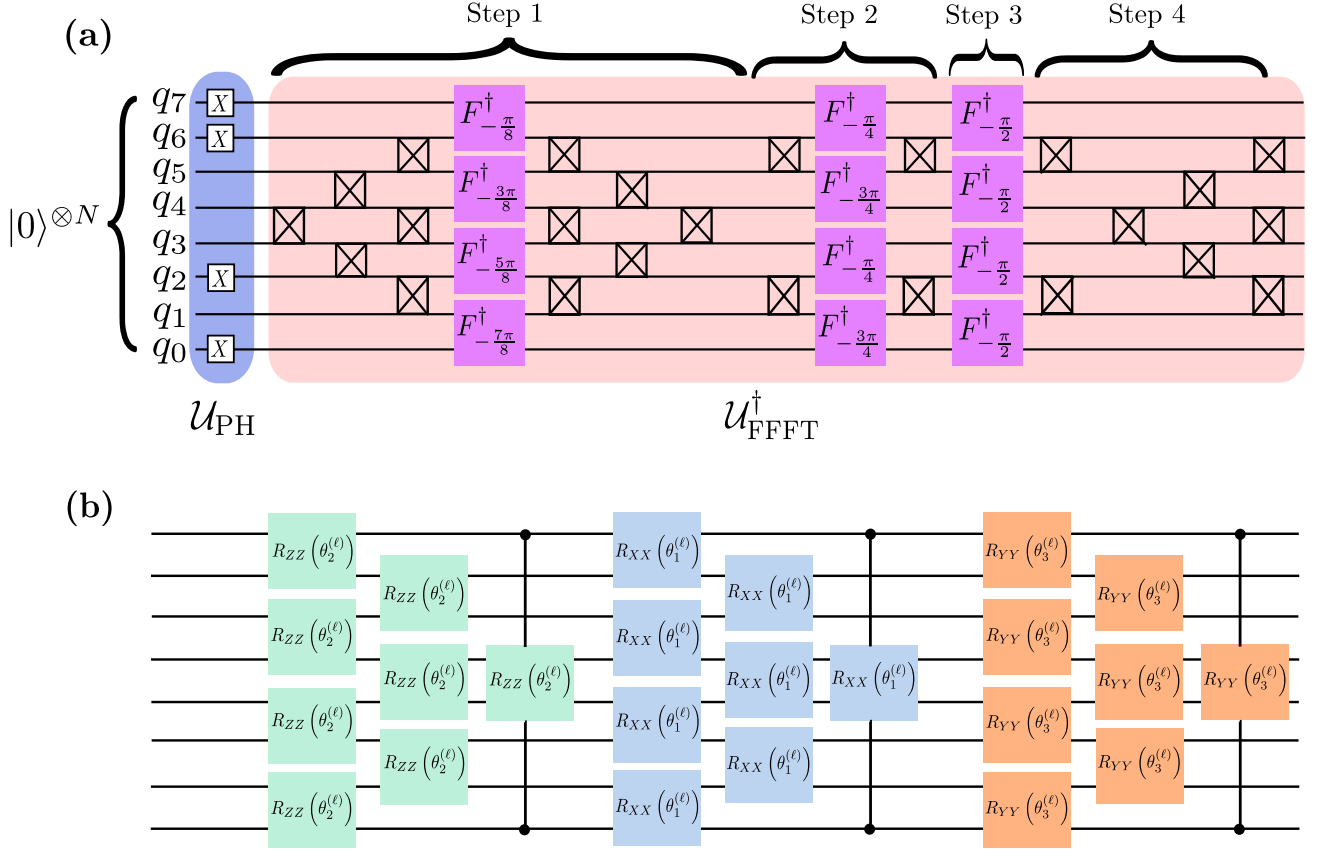}
\caption{Quantum circuit diagrams of the variational ansatz for quasiparticle simulation for an $N=8$ qubit system; convention for indexing the qubit register is shown in (a). (a) The free-fermion state preparation circuit $\mathcal{U}_{\text{PH}}$ and inverse fermionic fast Fourier transform (FFFT) circuit $\mathcal{U}_{\text{FFFT}}^{\dagger}$ are illustrated for the XX model. $\mathcal{U}_{\text{PH}}$ prepares a particle-hole symmetric state in the half-filled ($\mathcal{N}_{F}=4$) sector with total momentum $q=\frac{\pi}{4}$, and $\mathcal{U}_{\text{FFFT}}^{\dagger}$ is the 8 qubit inverse fermionic fast Fourier transform (FFFT) circuit. fSWAP gates are denoted by the crossed-out square boxes. (b) Illustration of a single layer of the HVA circuit, consisting of parameterized $R_{\text{XX}}$, $R_{\text{YY}}$, and $R_{\text{ZZ}}$ gates.} 
\label{fig:XXZansatz}
\end{figure*}

Having recast the XXZ Hamiltonian in the fermionic basis, we use the eigenstates of the XX Hamiltonian (which are free-fermion states) which serve as the initial states of the ansatz outlined in Sec.~\ref{sec:overview}. Because of the translation and global $U(1)$ symmetries of the XX Hamiltonian, each eigenstate of the XX Hamiltonian can be labeled by the quantum numbers $q$ and $\mathcal{N}_{F}$, where $q$ is the total momentum of the eigenstate and $\mathcal{N}_{F}$ labels the fermion number sector. The initial state preparation circuit for the XX model is illustrated in Fig.~\ref{fig:XXZansatz}(a), which consists of both a free-fermion state preparation circuit $\mathcal{U}_{\text{PH}}$ and an inverse Fourier transform circuit $\mathcal{U}_{\text{FFFT}}^{\dagger}$. The circuit $\mathcal{U}_{\text{PH}}$ prepares an eigenstate of $\mathcal{H}_{\text{XX}}$ in a sector specified by $\left(q,\mathcal{N}_{F}\right)$. The input state to $\mathcal{U}_{\text{PH}}$ is the qubit register initialized to be all in the $|0\rangle$ state, corresponding to the fermionic vacuum. The output state of $\mathcal{U}_{\text{PH}}$ is a free-fermion state in momentum space with fixed $\mathcal{N}_{F}$. The detailed and complete algorithm on the construction of $\mathcal{U}_{\text{PH}}$ to output a free-fermion state in any $\left(q,\mathcal{N}_{F}\right)$ sector is provided in Appendix~\ref{sec:XXmodelprep}. Here, we briefly summarize how to construct $\mathcal{U}_{\text{PH}}$ to output any free-fermion state in the half-filled sector $\mathcal{N}_{F}=N/2$. We adopt the convention that a computational basis state is expressed as $|q_{0},q_{1},\ldots,q_{N-1}\rangle$, where $q_{j}$ denotes the state of the $j^{\text{th}}$ qubit.
\begin{itemize}
\item\textit{Fermion-parity aware momentum grid:} As shown in Eq.~\eqref{eq:XXZboundary}, because the fermion parity $\mathcal{P}$ fixes the boundary condition, one must determine the allowed momentum values (i.e., the momentum grid) that is commensurate with the boundary condition. This means for even fermion parity, $k\in\left\{-\frac{\pi}{N}(N-1),\ldots,\frac{\pi}{N}(N-1)\right\}$ and for odd fermion parity, $k\in\left\{-\frac{\pi}{N}(N-2),\ldots,\pi\right\}$, where the momentum spacing of the grid is given by $\frac{2\pi}{N}$. The momentum grid values can be mapped to the indices of the qubit register via a \textit{parity-aware maps} $k_{\mathcal{P}}(m)$ given as
\begin{equation}\begin{split}\label{eq:fermionparityaware}
&k_{\mathcal{P}=+1}(m)=\frac{\pi}{N}(2m-(N-1))\\
&k_{\mathcal{P}=-1}(m)=\frac{\pi}{N}(2m-(N-2)),
\end{split}\end{equation} 
where $m\in\left\{0,\ldots,N-1\right\}$ are the indices of the qubit register. 
\item\textit{Construct a Single Particle-Hole (PH) State for $\mathcal{N}_{F}=N/2$ and fixed $q$:} Given the parity-aware momentum map $k_{\mathcal{P}}(m)$ from qubit index $m\in\mathcal{M}$ to momentum, choose an occupied state with momentum $k_{h}$ where a hole is to be created, and an unoccupied state with momentum $k_{p}$ where a particle is to be created, such that $q\equiv k_{p}-k_{h}\hspace{0.05cm}(\text{mod}\hspace{0.05cm}2\pi)$, and let $m_{h}$, $m_{p}$ satisfy $k_{\mathcal{P}}(m_{h})=k_{h}$, $k_{\mathcal{P}}(m_{p})=k_{p}$. With the qubit register initialized to $|0\rangle^{\otimes N}$, prepare the PH state in one pass by applying $X$ gates to all qubits corresponding to the half-filled Fermi sea \textit{except} $q_{m_{h}}$, and additionally applying an $X$ gate to $q_{m_{p}}$:
\begin{equation}\begin{split}
&\mathcal{U}_{\text{PH}}|0\rangle^{\otimes N}=\left(\prod\limits_{\mathcal{M}_{\text{FS}}\setminus\{m_{h}\}}X_{m}\right)X_{m_{p}}|0\rangle^{\otimes N},
\end{split}\end{equation}
where $\mathcal{M}_{\text{FS}}=\left\{0,\ldots,\frac{N}{4}-1\right\}\cup\left\{\frac{3N}{4},\ldots,N-1\right\}$ denotes set of qubit indices corresponding to the set of occupied momenta at half filling (i.e., the Fermi sea). As an example, for $N=8$ qubits, this set is $\mathcal{M}_{\text{FS}}=\left\{0,1\right\}\cup\left\{6,7\right\}$. For the initial PH state shown in Fig.~\ref{fig:XXZansatz}, the hole index is $m_{h}=1$ and the particle index is $m_{p}=2$, corresponding to momenta $k_{h}=-\frac{5\pi}{8}$ and $k_{p}=-\frac{3\pi}{8}$ respectively, according to the fermion-parity aware map given by Eq.~\eqref{eq:fermionparityaware}. This choice of particle and hole momenta results in a $q=\frac{\pi}{4}$ PH state in the $\mathcal{N}_{F}=4$ sector.
\end{itemize}
We note that for a given $q$, the momenta $k_p$ and $k_h$ are not unique, corresponding to the fact that particle-hole energy spectra span a finite width at each $q$. As explained earlier, for small $q$ this bandwidth is small, and the choice of $k_p$ and $k_h$ does not affect the variational result for $E(q)$. However, to approach the true band minimum at fixed $q$, the PH symmetry can be invoked to restrict the variational space and thereby target the lowest-energy configuration more efficiently. The details of this are discussed in Appendix~\ref{sec:XXmodelprep}.

After preparing the state with $\mathcal{U}_{\text{PH}}$, we apply the inverse Fourier transform $\mathcal{U}_{\text{FFFT}}^{\dagger}$ to convert to real space. As a unitary transformation, $\mathcal{U}_{\text{FFFT}}$ maps a Fock state in real-space to momentum-space, which implies
\begin{equation}\begin{split}\label{eq:FFFTunitary}
\mathcal{U}_{\text{FFFT}}c_{n}^{\dagger}\mathcal{U}_{\text{FFFT}}^{\dagger}=\frac{1}{\sqrt{N}}\sum\limits_{n}c_{n}^{\dagger}e^{ikn}\equiv\tilde{c}_{k}^{\dagger}.
\end{split}\end{equation}
We implement this unitary transformation with the fermionic fast Fourier transform (FFFT) algorithm~\cite{Verstraete2009,Ferris2014}, which follows a Cooley-Tukey-style~\cite{Cooley1965} radix-2 decomposition and achieves a $\text{O}\left(N\,\text{log}\,N\right)$ gate count where each gate operates on pairs of qubits; accordingly, this requires the number of qubits $N$ to be a power of $2$. This algorithm involves applying two types of gates, the fermion SWAP (fSWAP) gates, which exchange fermionic modes to proper ordering, and the two-qubit gates $F_{k}$, which assigns the relative phase between even/odd qubits as per the usual FFT algorithm. The fSWAP operator for two fermions can be expressed in the momentum-space fermionic basis as 
\begin{equation}\begin{split}\label{eq:Fnk}
\text{fSWAP}=\begin{pmatrix} 1 & 0 & 0 & 0 \\ 0 & 0 & 1& 0 \\ 0 & 1 & 0 & 0 \\ 0 & 0 & 0 & -1 \end{pmatrix}.
\end{split}\end{equation}
with the basis states ordered as $\{|00\rangle,|01\rangle,|10\rangle,|11\rangle\}$. The fSWAP operator acts on this basis as
\begin{equation}\begin{split}\label{eq:fSWAPII}
&\text{fSWAP}|00\rangle=|00\rangle,\\
&\text{fSWAP}|01\rangle=|10\rangle,\\
&\text{fSWAP}|10\rangle=|01\rangle,\\
&\text{fSWAP}|11\rangle=-|11\rangle,
\end{split}\end{equation}
as it should due to fermionic statistics. Additionally, there are a number of two-qubit $F_k$ gates that assign the phases appearing in the Fourier transform,
\begin{equation}\begin{split}\label{eq:Fnk}
F_k=\begin{pmatrix} 1 & 0 & 0 & 0 \\ 0 & \frac{1}{\sqrt{2}} & \frac{1}{\sqrt{2}}e^{ik}& 0 \\ 0 & \frac{1}{\sqrt{2}} & -\frac{1}{\sqrt{2}}e^{ik} & 0 \\ 0 & 0 & 0 & -e^{ik} \end{pmatrix}.
\end{split}\end{equation}
The quantum circuit implementation of the inverse FFFT, explained in detail in Refs.~\cite{Verstraete2009,Ferris2014} is shown in Fig.~\ref{fig:XXZansatz} (a) for $N=8$ qubits. Although the circuit is rather complicated, intuitively it can be understood as folding the Brillouin zone (BZ) in half recursively [Steps 1,2,3 in Fig.~\ref{fig:XXZansatz} (a) ]. At every step, this is done by pairing up momenta $k$ and $k+\pi$ (the size of the BZ is rescaled to be $2\pi$ at every step). This increases the unit cell size (as well as the number of bands) by a factor of two, and the $F^\dagger_k$ gates convert the Bloch states from the band basis to the sublattice basis. For $N=2^m$, after $m$ steps, the entire system consists of just one unit cell, and the resulting wavefunction becomes a Fock state in real space. Lastly [Step (4) in Fig.~\ref{fig:XXZansatz} (a)], we rearrange the qubits using fSWAP's to the correct real-space order.

\subsection{Implementing the HVA}\label{subsec:hva}

Next, we outline the construction of the parameterized HVA circuit. Because the input state lies in a fixed $\left(q,\mathcal{N}_{F}\right)$ sector, the ideal ansatz would preserve translation, $U(1)$, and $\mathbb{Z}_{2}$ symmetries so that $|\Psi(\boldsymbol{\theta})\rangle$ remains in the same sector throughout optimization. In practice, enforcing all symmetries while maintaining shallow circuit depth is challenging. Below we compare different HVA implementations, outline their trade-offs in depth, parameterization, and which symmetries they preserve, and justify the choice used in this work.

First, we consider the following implementation of the HVA,
\begin{equation}\begin{split}\label{eq:HVAI}
&\mathcal{U}_{\text{HVA}}(\boldsymbol{\theta})=\prod\limits_{\ell=1}^{p}\left[\left(\prod\limits_{m=0}^{N-1}R_{\text{YY},m}\left(\theta_{3}^{(\ell)}\right)\right)\left(\prod\limits_{n=0}^{N-1}R_{\text{XX},n}\left(\theta_{1}^{(\ell)}\right)\right)\right.\\
&\left.\left(\prod\limits_{j=0}^{N-1}R_{\text{ZZ},j}\left(\theta_{2}^{(\ell)}\right)\right)\right],
\end{split}\end{equation}
where $R_{\alpha\alpha,n}\left(\theta\right)=\exp\left\{-\frac{i\theta}{2}\sigma_{n}^{\alpha}\sigma_{n+1}^{\alpha}\right\}$ for $\alpha\in\left\{x,y,z\right\}$ and $n\hspace{0.1cm}(\text{mod}\hspace{0.1cm}N)$ is the index of the lattice site. The quantum circuit diagram for a single layer of this HVA is illustrated in Fig.~\ref{fig:XXZansatz}(b), and the total number of parameters in this ansatz is $3p$, where $p$ is the number of layers. This implementation of the Trotter decomposition is highly symmetric, preserving the translation, global $\mathbb{Z}_{2}$ parity, and spatial inversion symmetries of the Hamiltonian exactly. A single layer of Eq.~\eqref{eq:HVAI} preserves translation symmetry because each block is a uniform product of identical two-site gates applied on every bond, implying that $\left[R_{\alpha\alpha,n}\left(\theta_{1}^{(\ell)}\right),R_{\alpha\alpha,n+1}\left(\theta_{1}^{(\ell)}\right)\right]=0$ for all $n$ and $\alpha\in\{x,y,z\}$. Denoting the one-site translation operator as $\mathcal{T}$, one has
\begin{equation}\begin{split}\label{eq:translationHVAI}
\mathcal{T}\left[\prod\limits_{n=0}^{N-1}R_{\alpha\alpha,n}(\boldsymbol{\theta})\right]\mathcal{T}^{\dagger}=\prod\limits_{n=0}^{N-1}R_{\alpha\alpha,n}(\boldsymbol{\theta})
\end{split}\end{equation}
which implies $\mathcal{T}\mathcal{U}_{\text{HVA}}(\boldsymbol{\theta})\mathcal{T}^{\dagger}=\mathcal{U}_{\text{HVA}}(\boldsymbol{\theta})$. However, it breaks $U(1)$ symmetry since $\left[S_{z},\sigma_{n}^{\alpha}\sigma_{n+1}^{\alpha}\right]\neq 0$ for $\alpha\in\{x,y\}$, so the separate rotations $R_{\text{XX}}$ and $R_{\text{YY}}$ do not conserve $S_{z}$ (or equivalently, $\mathcal{N}_{F}$).

We will use the ansatz in Eq.~\eqref{eq:HVAI} in this work. Due to its translation and parity invariance, the VQE optimization remains confined to the desired $(q,\mathcal{P})$ sector. Although the separate $R_{\text{XX}}$ and $R_{\text{YY}}$ blocks do not conserve $\mathcal{N}_{F}$ and thus break $U(1)$, mixing into sectors with $\Delta\mathcal{N}_{F}=\pm 2,\pm 4,\ldots$ by adding even numbers of quasiparticles/holes is energetically disfavored; empirically, the optimized states remain (to high accuracy) in the intended half-filled sector and yield the correct lowest excitation at each $q$, as we will show in Sec.~\ref{sec:results}. 
As Eq.~\eqref{eq:HVAIII} shows, one can restore the one-site translation symmetry while keeping $U(1)$ by conjugating the block containing the momentum-dependent phases with FFFT/Givens rotations, but this inserts two nontrivial unitary transformations per layer and dramatically increases the gate count. Given our goal of resolving the lowest excitation energy at each momentum $q$, the symmetry/depth trade-off favors Eq.~\eqref{eq:HVAI}; it preserves the symmetries most critical for targeting the lowest energy $\left(q,\mathcal{P}\right)$ sectors and remains compact enough to train reliably. 

By contrast, another possible implementation of the HVA that exactly preserves both the $U(1)$ and $\mathbb{Z}_{2}$ parity symmetries is the following,
\begin{equation}\begin{split}\label{eq:HVAII}
&\tilde{\mathcal{U}}_{\text{HVA}}(\boldsymbol{\theta})=\prod\limits_{\ell=1}^{p}\left[\left(\prod\limits_{n=0}^{N-1}R_{\text{XX}+\text{YY},n}\left(\theta_{1}^{(\ell)}\right)\right)\right.\\
&\left.\left(\prod\limits_{j=0}^{N-1}R_{\text{ZZ},j}\left(\theta_{2}^{(\ell)}\right)\right)\right],
\end{split}\end{equation}
where we have introduced the following two-qubit rotation gate,
\begin{equation}\begin{split}
&R_{\text{XX}+\text{YY},n}\left(\theta\right)=\exp\left\{-\frac{i}{2}\theta\left(\sigma_{n}^{x}\sigma_{n+1}^{x}+\sigma_{n}^{y}\sigma_{n+1}^{y}\right)\right\}.
\end{split}\end{equation}
This ansatz separately groups $R_{\text{XX}+\text{YY},n}\left(\theta_{1}^{(\ell)}\right)$ terms and $R_{\text{ZZ},n}\left(\theta_{2}^{(\ell)}\right)$ terms on each bond within their own dedicated blocks. This arrangement preserves $U(1)$ symmetry because $\left[S_{z},\sigma_{n}^{x}\sigma_{n+1}^{x}+\sigma_{n}^{y}\sigma_{n+1}^{y}\right]=0$. Compared to Eq.~\eqref{eq:HVAI}, this ansatz contains $2p$ parameters. Although this ansatz does closely resemble the decomposition discussed in Sec.~\ref{sec:overview}, and has been shown to converge to the ground state of interacting systems~\cite{AnselmeMartin2022,Falide2025} such as the Fermi-Hubbard model, it fails to preserve translation symmetry. This is because under the Trotter decomposition, $\left[R_{\text{XX}+\text{YY},n}\left(\theta_{1}^{(\ell)}\right),R_{\text{XX}+\text{YY},n+1}\left(\theta_{1}^{(\ell)}\right)\right]\neq 0$, which implies that $\left[\mathcal{T},\prod\limits_{n=0}^{N-1}R_{\text{XX}+\text{YY},n}\left(\theta_{1}^{(\ell)}\right)\right]\neq 0$. For performing VQE for quasiparticle simulation, this ansatz is inadequate, as the outcome is no longer a momentum eigenstate and will get mixed with the ground state with $q=0$. 

We note in passing that it is possible to preserve both the one-site translation and $U(1)$ symmetries using the following HVA proposed in Ref.~\cite{Babbush2018} (the details of this HVA construction are provided in Appendix~\ref{sec:alternateHVA}),
\begin{equation}\begin{split}\label{eq:HVAIII}
&\bar{\mathcal{U}}_{\text{HVA}}(\boldsymbol{\theta})=\prod\limits_{\ell=1}^{p}\left[\left(\mathcal{U}_{\text{FFFT}}^{\dagger}\left(\prod\limits_{m=0}^{N-1}R_{\text{Z},m}\left(\theta_{1,m}^{(\ell)}\right)\right)\mathcal{U}_{\text{FFFT}}\right)\right.\\
&\left.\left(\prod\limits_{j=0}^{N-1}R_{\text{Z},j}\left(\theta_{2}^{(\ell)}\right)\right)\left(\prod\limits_{n=0}^{N-1}R_{\text{ZZ},n}\left(\theta_{3}^{(\ell)}\right)\right)\right].
\end{split}\end{equation}
In the above, we have introduced the notation for the single-qubit rotation gate $R_{\text{Z},m}\left(\theta\right)=\exp\left\{-\frac{i\theta}{2}\sigma_{m}^{z}\right\}$. Eq.~\eqref{eq:HVAIII} generally contains $(N+2)p$ parameters because each term in the conjugated FFFT block can be assigned a distinct parameter. However, this ansatz involves multiple applications of FFFT's and is not as efficient as Eq.~\eqref{eq:HVAI} for our purposes. 

\section{Results}\label{sec:results} 

\begin{figure}
\includegraphics[width=0.5\textwidth]{"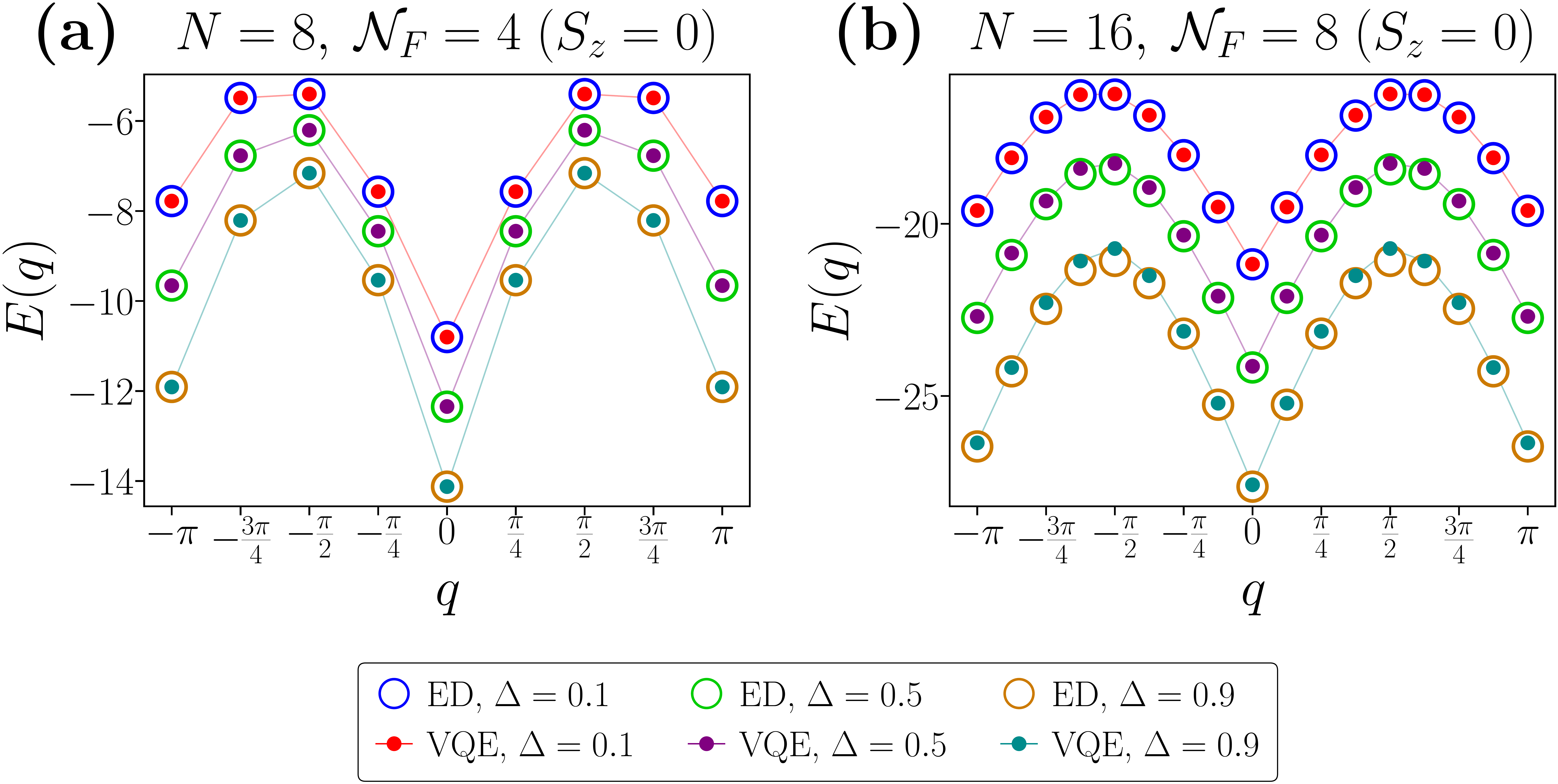"}
\caption{Plots of the excitation spectrum for different values of the anisotropy strength $\Delta$ obtained from VQE and exact diagonalization (ED). The quasiparticle excitation spectrum is plotted for the half-filled sector ($S_{z}=0$) for (a) $N=8$ and (b) $N=16$. For reference, the ground state of the XXZ model is located in the half-filled sector at momentum $q=0$. The solid circles are the VQE-optimized energy values, and the open circles are the ED values.}
\label{fig:quasiparticle}
\end{figure}

\begin{figure}
\includegraphics[width=0.5\textwidth]{"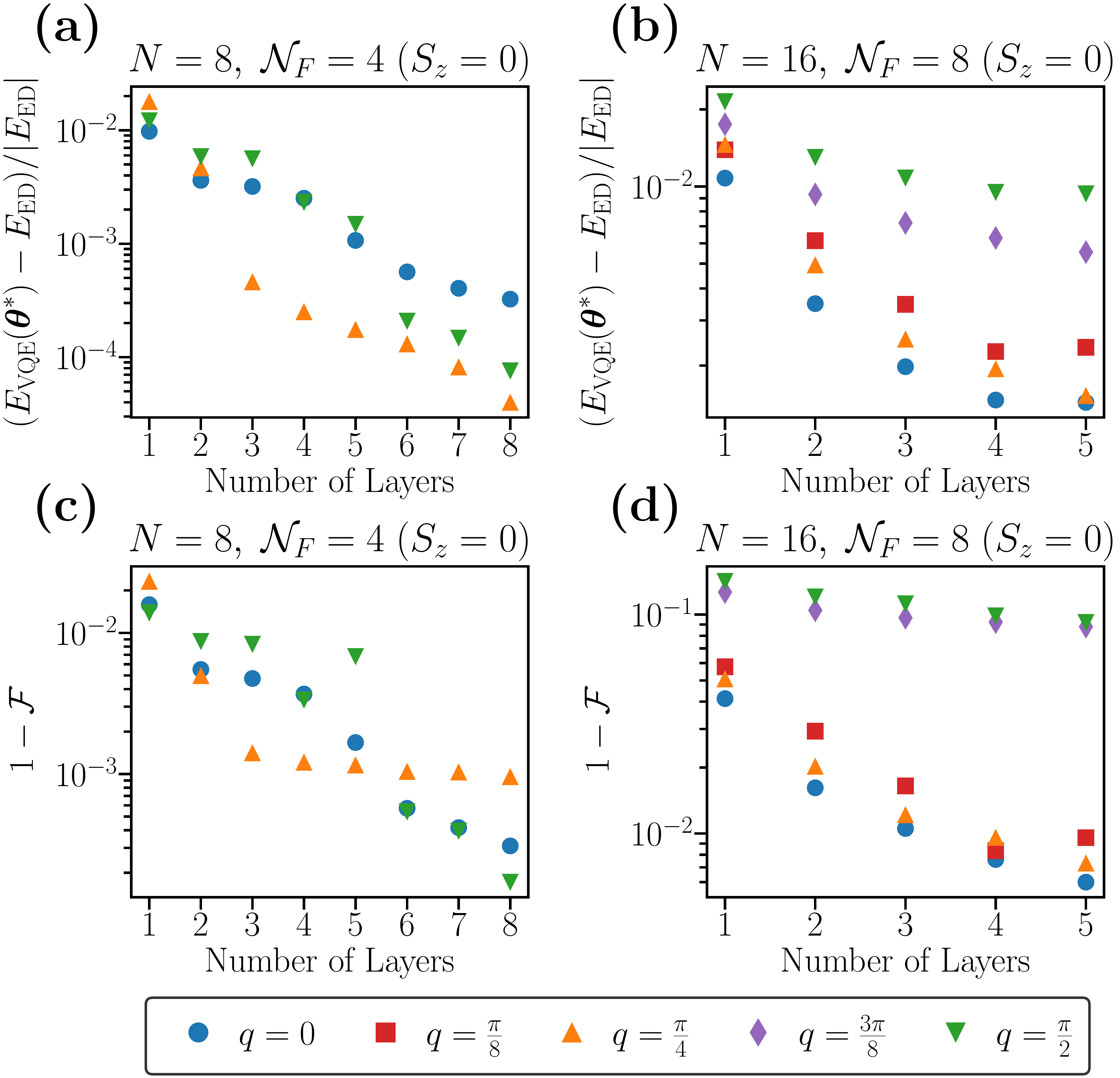"}
\caption{Semi-log plots of the obtained from VQE for the optimized energies relative errors ((a)-(b)) and the infidelities ((c)-(d)) for $\Delta=0.5$ in the $\mathcal{N}_{F}=N/2$ sector. The errors are plotted for the lowest momentum values ($q\in\left\{0,\frac{\pi}{4},\frac{\pi}{2}\right\}$ for $N=8$  in (a) \& (c) and $q\in\left\{0,\frac{\pi}{8},\frac{\pi}{4},\frac{3\pi}{8},\frac{\pi}{2}\right\}$ for $N=16$ in (b) \& (d)).}
\label{fig:errors}
\end{figure}

We demonstrate the capabilities of quasiparticle VQE for the XXZ Hamiltonian for $N=8$ and $N=16$ qubits, with periodic boundary conditions present. The main results are shown in Figs.~\ref{fig:quasiparticle} \& \ref{fig:errors} and Table~\ref{tab:velocities}. The specific computational details related to VQE and Figs.~\ref{fig:quasiparticle} \& \ref{fig:errors} and Table~\ref{tab:velocities} are provided in Appendix~\ref{sec:methodologicaldetails}. Throughout this section, we compare the VQE-optimized eigenstates and energies to the eigenstates and eigenvalues reported by exact diagonalization (ED) in order to assess accuracy. As a reference for where these states are located in the spectrum of the XXZ model, we provide a table of the target ED eigenvalue indices for each $\left(q,\mathcal{N}_{F}\right)$ subspace for both $N=8$ and $N=16$ qubits in Appendix~\ref{sec:tables}. Although we focus on the half-filled sector in this section, we also performed VQE to obtain the energies in the one-hole doped sector $\mathcal{N}_{F}=N/2-1$ which has fermion parity $\mathcal{P}=-1$, the results of which are provided in Appendix~\ref{sec:additionalresults}.

\begin{table}
\centering
\caption{Renormalized velocity of the bosonic quasiparticle (two-spinon excitation) obtained from VQE and Bethe ansatz (given by Eq.~\eqref{eq:theoreticalvelocity}).}
\label{tab:velocities}
\begin{tabular}{cccc}
\hline
$\Delta$ & VQE ($N=8$) & VQE ($N=16$) & Theory \\
\hline
$0.1$ & 4.113472 & 4.223552 & 4.251032 \\
$0.5$ & 4.958596 & 5.183916 & 5.196152 \\
$0.9$ & 5.839341 & 6.029963 & 6.072315 \\
\hline
\end{tabular}
\end{table}

Fig.~\ref{fig:quasiparticle} shows the quasiparticle dispersion obtained from both VQE and ED. For $N=8$ qubits, the results obtained from VQE are in excellent agreement with the results obtained from ED, across all momentum values and anisotropy strengths. Similar results are observed for $N=16$ qubits. The deviations between the VQE energy estimates and ED energies are larger at higher values of $\Delta$. This is because as $\Delta$ is increased, interactions play a larger role by inducing correlations between the fermionic excitations, and the HVA becomes less accurate. Fig.~\ref{fig:quasiparticle} captures the lowest energies of the quasiparticle excitation spectrum, which matches the bottom bounding curve highlighted in red in the particle-hole excitation spectrum of the model shown in Fig.~\ref{fig:XXSpectra} (b). Although the excitation energies are plotted over the full momentum range $q\in\{-\pi,\ldots\pi\}$ in Fig.~\ref{fig:quasiparticle}, only the excitations near $q=0,\pi$ correspond to long-lived quasiparticles. The renormalized velocity $v$ of the low-lying excitations can be estimated from the quasiparticle dispersion shown in Fig.~\ref{fig:quasiparticle} by taking the slope of the line between the ground state energy $E(0)$ and the excited state energy $E\left(2\pi/N\right)$ in the $\mathcal{N}_{F}=N/2$ sector. 
 In Table~\ref{tab:velocities}, we compare the renormalized velocity $v$ for quasiparticles ($q\to 0$ limit) obtained from the VQE energy estimates with the theoretical value established in Eq.~\eqref{eq:theoreticalvelocity}. The optimal VQE estimates for the renormalized velocity are largely consistent with the theoretical values established by Bethe ansatz and improve dramatically as the system size is increased. Thus, the results for the renormalized velocity $v$ obtained from VQE confirms the linearly-dispersing nature of the bosonic quasiparticles of the Luttinger phase. 

To determine how the accuracy of VQE is impacted by the number of HVA layers, we plot the relative energy errors and the infidelities (fidelity errors) in Fig.~\ref{fig:errors} for $\Delta=0.5$ and a subset of momentum values $q$. The relative errors of the optimized energies are determined by calculating
$(E_{\text{VQE}}(\boldsymbol{\theta}^*)-E_{\text{ED}})/\left|E_{\text{ED}}\right|$. Here, $E_{\text{VQE}}(\boldsymbol{\theta}^*)$ is the optimized energy corresponding to the VQE-optimized state $|\Psi(\boldsymbol{\theta}^*)\rangle$ with fixed momentum $q$ in the half-filled sector, and $E_{\text{ED}}$ is the corresponding eigenvalue of the $\left(q,\mathcal{N}_{F}\right)$ subspace obtained from ED. The infidelities are calculated using $1-\mathcal{F}$ where the fidelity is given by
$\mathcal{F}=\left|\langle\Psi(\boldsymbol{\theta}^*)|\Psi_{\text{ED}}\rangle\right|^{2}$. For both $N=8$ and $N=16$ qubits, increasing the number of layers reduces both the relative energy error and infidelity of the VQE-optimized state. For $N=8$ qubits shown in Figs.~\ref{fig:errors} (a) \& (c), each circuit depth $p\geq 2$ yields a VQE-optimized state with relative energy error and infidelity less than $1\%$ for all $q\in\left\{0,\frac{\pi}{4},\frac{\pi}{2}\right\}$. To obtain a better understanding on how circuit depth impacts accuracy, we turn our attention to larger system sizes such as $N=16$ qubits in Figs.~\ref{fig:errors} (b) \& (d). This shows a clear distinction between each of the momentum values in how the circuit depth affects the relative energy errors and infidelities of the VQE-optimized state. The ground state ($q=0$) has the lowest relative energy error and infidelity and rapidly decreases as number of HVA layers is increased. For the excited states with momentum at $q=\frac{3\pi}{8}$ and $q=\frac{\pi}{2}$, the relative energy errors and infidelities are roughly an order of magnitude higher compared to those of the ground state ($q=0$), and decrease more slowly with circuit depth. This is expected as at these momenta, the excitation no longer corresponds to a quasiparticle with a sharp spectral function, and VQE performs worse.

The decreasing trend in the errors observed in Figs.~\ref{fig:errors} (b) \& (d) indicates that a significant increase in the number of HVA layers is required to achieve the same level of accuracy for the excited states compared to that of the ground state. These features are typical of the HVA and are consistent with recent studies exploring low-lying excited states in interacting spin models using the HVA~\cite{Chen2025}.

\section{Conclusion}
\label{sec:conclusion}

In this paper we have developed a momentum-resolved, symmetry-preserving VQE scheme to simulate quasiparticle excitations in strongly correlated systems. The variational ansatz initializes a free-fermion particle-hole state at fixed momentum $q$ (prepared via an inverse FFFT) which is varied by a Hamiltonian Variational Ansatz (HVA) that preserves translation and $\mathbb{Z}_{2}$ parity symmetries while optimizing towards an excited eigenstate of the interacting model. While the HVA does not preserve the U(1) symmetry explicitly, energetic optimization effectively ensures the correct U(1) quantum number.  Applied to the spin-1/2 XXZ chain, this VQE framework reproduces the dispersion of the bosonic quasiparticles accurately across different anisotropy strengths when compared to results obtained from Bethe ansatz. These results demonstrate that a targeted VQE approach incorporating symmetries into the ansatz construction can yield several of the low-lying excited states in interacting quantum many-body systems. Additionally, our approach also demonstrates the advantage of fermionizing spin models such as the XXZ chain to prepare its quasiparticle states.

The approach examined in this work can also be straightforwardly applied to simulate quasiparticles in fermionic systems such as the Fermi-Hubbard model. A natural direction for future study would be to simulate quasiparticle excitations in other occupation number sectors, using the HVA given by Eq.~\eqref{eq:HVAIII} instead of Eq.~\eqref{eq:HVAI}, as this preserves all symmetries, including $U(1)$. Assessing the trade-off in circuit depth versus accuracy of these two ans\"atze is a pertinent direction of study. Another direction for future work is improving the accuracy of the VQE-optimized excited states by combining the framework discussed here with methods such as subspace-search VQE~\cite{Nakanishi2019} to more effectively target the higher total momentum $q$ subspaces. Furthermore, although we focused on noiseless optimization in this work (see Appendix~\ref{sec:methodologicaldetails}), an important direction for future study is examining the impact of noise (e.g., statistical noise from shot-based measurements) on optimization performance to assess the feasibility of hardware implementation, specifically for the quantum simulation of excited states.

As a closing perspective, these results motivate broader exploration of quantum simulation of excited states in strongly correlated systems via symmetry-aware ans\"atze and tailored initial state preparation. The XXZ model, which is integrable and can be solved analytically, serves here as a proof of concept for our quasiparticle VQE framework: we initialize exact eigenstates of the XX model and use them to target quasiparticle excitations of the interacting XXZ chain. A natural extension is to consider Hamiltonians of the form $\mathcal{H}=\mathcal{H}_{\text{QP}}+\Delta\mathcal{H}'$, where $\mathcal{H}_{\text{QP}}$ is diagonal in a quasiparticle basis and $\mathcal{H}'$ introduces quasiparticle mixing/interactions, which may not necessarily be integrable. In this setting, one may employ an ansatz composed of (i) an eigenstate preparation circuit for $\mathcal{H}_{\text{QP}}$ that maps a quasiparticle eigenstate in momentum-space to real-space and (ii) an HVA circuit with Trotterized layers alternating parameterized unitaries $\exp\left\{-i\theta_{1}^{(\ell)}\mathcal{H}_{\text{QP}}\right\}$ and $\exp\left\{-i\theta_{2}^{(\ell)}\mathcal{H}'\right\}$ for each layer $\ell\in\left\{1,\ldots,p\right\}$. Many exactly solvable models provide natural choices for $\mathcal{H}_{\text{QP}}$\textemdash e.g., the XY chain~\cite{Lieb1961}, the transverse field Ising model~\cite{Pfeuty1970}, and the Kitaev honeycomb model~\cite{Kitaev2006}\textemdash for which explicit eigenstate circuits are known~\cite{Schmoll2017,CerveraLierta2018,Xiao2021,Farreras2025,Kokcu2025}. A further direction is to search for quasiparticle excitations in strongly correlated systems with longer-range interactions and hoppings that preserve translation symmetry but generically break integrability, making them well suited for this approach. Taken together, these considerations highlight that a momentum-space, symmetry-adapted VQE framework based on partitioning the Hamiltonian into free and interacting components can be systematically extended to a broad class of models.\\

\section*{Data Availability}
The codes and data used for all VQE calculations and plots shown in this article have been made publicly available~\cite{GitHub2025}.

\begin{acknowledgments}
We thank Shuyi Li and Chunjing Jia for useful discussions. S.V. is supported by the Dirac Postdoctoral Fellowship, sponsored by the National High Magnetic Field Laboratory (NHMFL). Y.W. is supported by NSF under grant number DMR-2045781. Numerical VQE calculations in this work were performed using the \texttt{qsimcirq}~\cite{qsimcirq2020}, OpenFermion~\cite{McClean2020}, and Py-BOBYQA~\cite{Cartis2018} libraries.
\end{acknowledgments}

\nocite{REVTEX42Control,apsrev42Control}
\bibliographystyle{apsrev4-2}
\bibliography{QPVQE}

\appendix

\section{Computational Details}\label{sec:methodologicaldetails}

All VQE calculations were performed using noiseless statevector simulations with the simulator provided in the \texttt{qsimcirq} package. The FFFT was implemented using functions provided in the OpenFermion~\cite{McClean2020} library. For both values of $N=8$ and $N=16$ qubits, VQE was performed separately for each momentum value $q\in\left\{-\pi+\frac{2\pi}{N},\ldots,\pi\right\}$ in the half-filled sector $\mathcal{N}_{F}=N/2$ (or equivalently, $S_{z}=0$ sector). The ansatz used for VQE is given by Eq.~\eqref{eq:finalansatz}, with the HVA given by Eq.~\eqref{eq:HVAI}. Classical optimization was performed using the gradient-free optimization algorithm BOBYQA~\cite{Powell2009} from the \texttt{Py-BOBYQA} package~\cite{Cartis2018}. The parameter vector $\boldsymbol{\theta}$ contains $3p$ parameters, where $p$ is the number of layers, and is expressed as $\boldsymbol{\theta}=\begin{pmatrix} \theta_{1}^{(1)}, & \theta_{2}^{(1)}, & \theta_{3}^{(1)}, & \cdots & \theta_{1}^{(p)}, & \theta_{2}^{(p)}, & \theta_{3}^{(p)} \end{pmatrix}$. 

For each run of VQE, we instantiate the ansatz $|\Psi(\boldsymbol{\theta})\rangle$ with $3p$ randomly selected parameters from the interval $[0,2\pi)$ and then perform optimization using BOBYQA. As a gradient-free optimizer, BOBYQA does not require the costly computation of gradients via finite-difference methods that other gradient-based optimizers require and can converge quickly. However, BOBYQA is susceptible to local extrema and its performance depends on the initial choice of parameters, which has been observed in related works~\cite{Li2023}. Therefore, to obtain the most optimal results from BOBYQA, we employ a multiple initial conditions strategy where each run is instantiated with randomized initial parameters, which we discuss below.

\begin{figure}
\centering
\includegraphics[width=0.49\textwidth]{"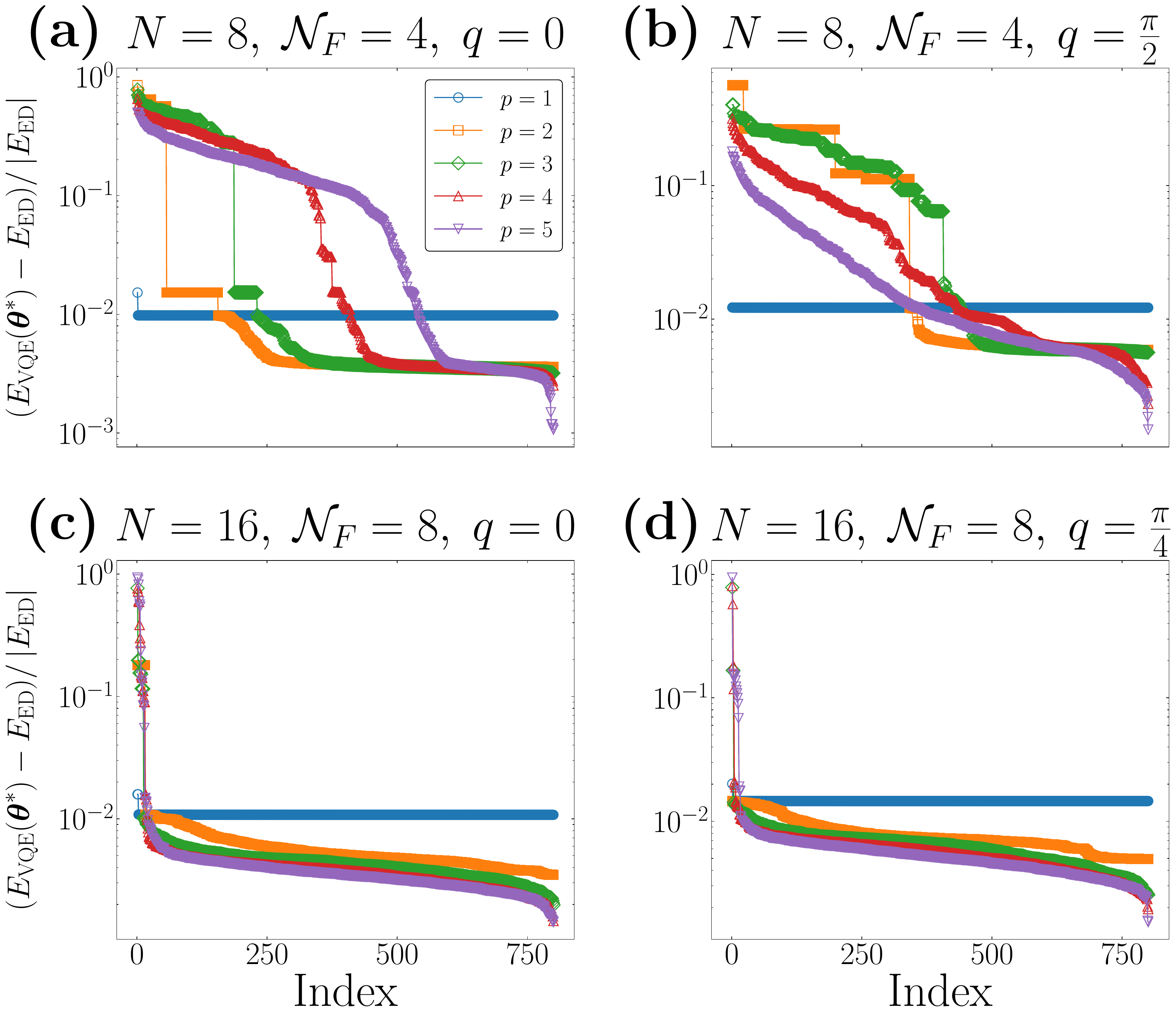"}
\caption{Semi-log distribution plots of the relative energy errors for various HVA circuit depths in the half-filled sector ($\mathcal{N}_{F}=N/2$ or $S_{z}=0$) for different values of $q$ and $\Delta=0.5$. The different circuit depths are denoted by $p$, which specifies the number of layers. The horizontal axis provides the index of each run.}
\label{fig:errordistributions}
\end{figure} 

The VQE-optimized energy values shown in Fig.~\ref{fig:quasiparticle} were obtained from an 8-layer HVA for $N=8$ qubits, and a 5-layer HVA for $N=16$ qubits. The values reported in both Figs.~\ref{fig:quasiparticle} \& \ref{fig:errors} were obtained using the following procedure. With the momentum $q$, fermion number $\mathcal{N}_{F}$, and circuit depth $p$ fixed, $800$ independent runs of VQE were performed with the BOBYQA optimizer. The relative energy error distributions of these runs is shown in Fig.~\ref{fig:errordistributions}. Then, the VQE-optimized state with the lowest relative energy error from the $800$ runs (i.e., the minimum of the distribution of relative energy errors) with respect to the target ED eigenvalue was plotted in Fig.~\ref{fig:errors} (a) \& (b) (and Fig.~\ref{fig:quasiparticle} (a) \& (b)). The corresponding infidelity for this selected VQE-optimized state at each circuit depth $p$ was also plotted in Fig.~\ref{fig:errors} (c) \& (d). VQE was performed using this procedure for each circuit depth $p\in\left\{1,\ldots,8\right\}$ at each momentum $q\in\left\{-\pi,-\frac{3\pi}{4},\ldots,\pi\right\}$ for each anisotropy strength $\Delta\in\left\{0.1,0.5,0.9\right\}$ for $N=8$ qubits. In order to save time on computation for the larger system size of $N=16$ qubits, this same procedure was applied only for the circuit depths $p\in\left\{1,2,3,4,5\right\}$ at each momentum $q\in\left\{-\pi,-\frac{7\pi}{8},\ldots,\pi\right\}$ for the anisotropy strengths $\Delta\in\left\{0.1,0.5\right\}$. For $\Delta=0.9$ and $N=16$ qubits, VQE runs were only performed with a $p=8$ layer HVA.

In Fig.~\ref{fig:quasiparticle} and Table~\ref{tab:velocities}, the VQE energy estimates reported for each momentum value $q\in\left\{-\pi,-\frac{3\pi}{4},\ldots,\pi\right\}$ and $\Delta\in\left\{0.1,0.5,0.9\right\}$ for the $N=8$ qubit system size was obtained from an 8-layer HVA. For the $N=16$ qubit system size, the VQE energy estimate plotted for each momentum value $q\in\left\{-\pi,-\frac{7\pi}{8},\ldots,\pi\right\}$ for $\Delta\in\left\{0.1,0.5\right\}$ was obtained from a 5-layer HVA. To ensure the best fit with ED results at the larger anisotropy strength of $\Delta=0.9$, the VQE energy estimates plotted were obtained from an 8-layer HVA.

Additionally, to obtain the VQE-optimized energies for each $|q|>\frac{2\pi}{N}$, we constructed $\mathcal{U}_{\text{PH}}$ so that it outputs a Bell pair/EPR PH-symmetric state according to the protocol detailed in Appendix~\ref{sec:XXmodelprep}.

\section{Numerical Results for the $\mathcal{N}_{F}=N/2-1$ Sector}\label{sec:additionalresults}

\begin{figure}[htbp]
\includegraphics[width=0.5\textwidth]{"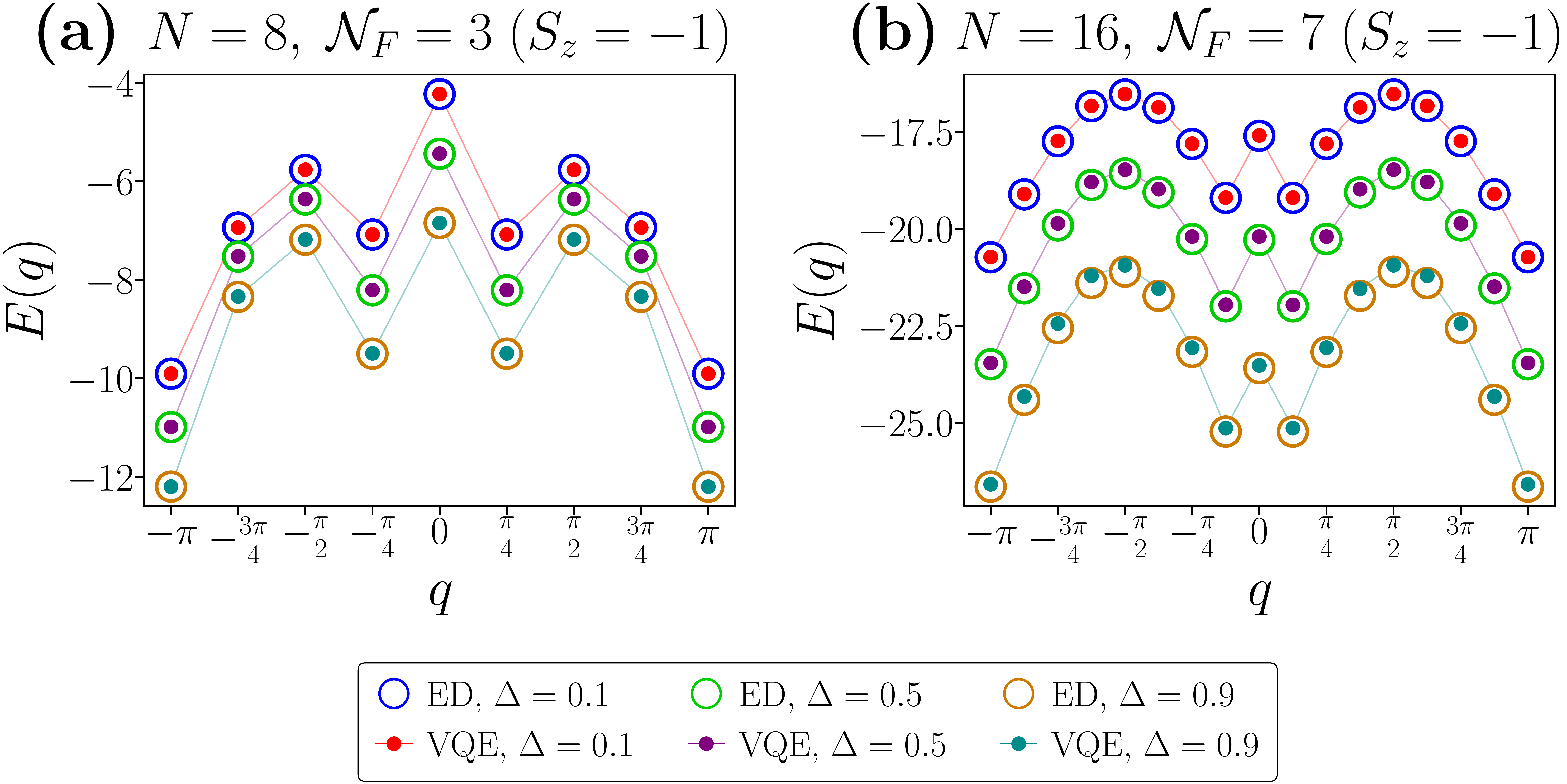"}
\caption{Plots of the excitation spectrum for different values of the anisotropy strength $\Delta$ obtained from VQE and exact diagonalization (ED). The quasiparticle excitation spectrum is plotted for the one-hole doped sector ($S_{z}=-1$) for (a) $N=8$ and (b) $N=16$. For reference, the lowest energy state in the $N/2-1$ sector is located at momentum $q=\pi$.}
\label{fig:quasiparticleapp}
\end{figure}

\begin{figure}[htbp]
\includegraphics[width=0.5\textwidth]{"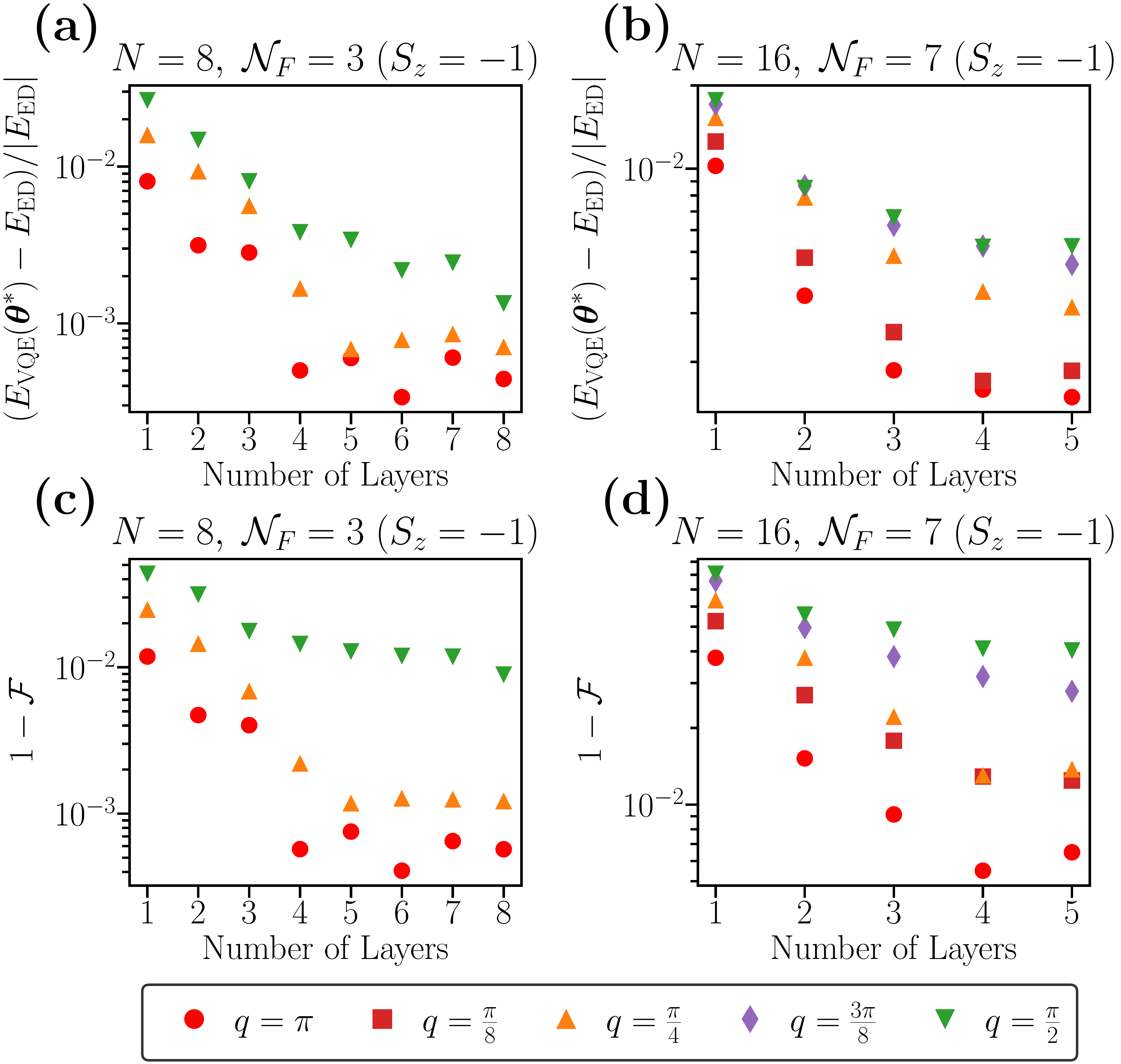"}
\caption{Semi-log plots of the obtained from VQE for the optimized energies relative errors ((a)-(b)) and the infidelities ((c)-(d)) for $\Delta=0.5$ in the $\mathcal{N}_{F}=N/2-1$ sector. The errors are plotted for the lowest momentum values ($q\in\left\{\frac{\pi}{4},\frac{\pi}{2},\pi\right\}$ for $N=8$  in (a) \& (c) and $q\in\left\{\frac{\pi}{8},\frac{\pi}{4},\frac{3\pi}{8},\frac{\pi}{2},\pi\right\}$ for $N=16$ in (b) \& (d)).}
\label{fig:errorsapp}
\end{figure}

\begin{figure}[htbp]
\centering
\includegraphics[width=0.49\textwidth]{"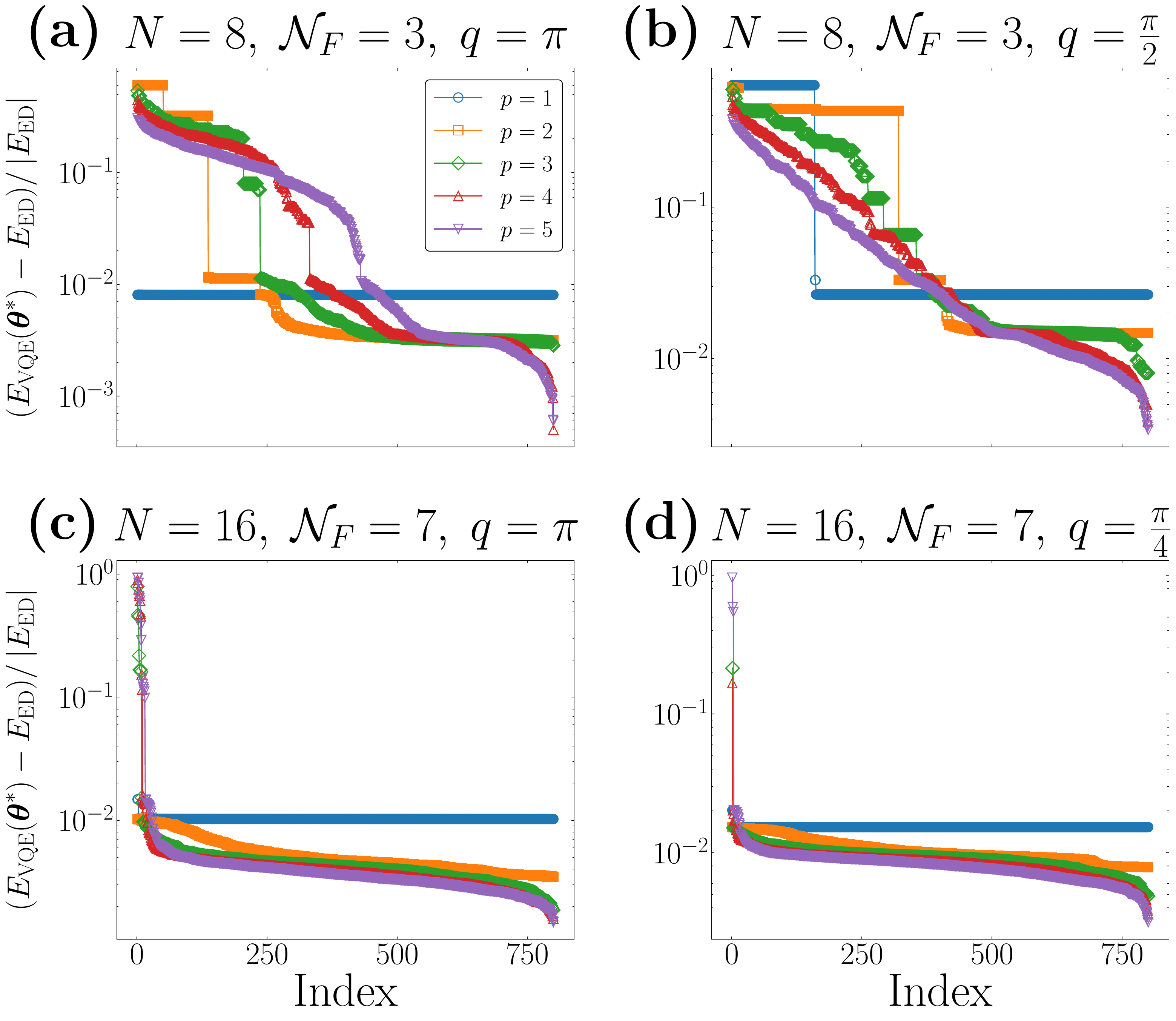"}
\caption{Semi-log distribution plots of the relative energy errors for various HVA circuit depths in the one-hole doped sector ($\mathcal{N}_{F}=\frac{N}{2}-1$ or $S_{z}=-1$) for different values of $q$ and $\Delta=0.5$. The different circuit depths are denoted by $D$, which specifies the number of layers. The horizontal axis provides the index of each run.}
\label{fig:errordistributionsapp}
\end{figure}

In this appendix, we provide additional numerical results for the ``one-hole doped" $\mathcal{N}_{F}=N/2-1$ fermion number sector ($S_{z}=-1$ sector) obtained using the HVA given by Eq.~\eqref{eq:HVAI}. The results relevant to this appendix are displayed in Figs.~\ref{fig:quasiparticleapp}-\ref{fig:errordistributionsapp}. 

\section{Jordan-Wigner Transformation}\label{sec:JWTransform}

In this appendix, we provide a brief review of the Jordan-Wigner transformation and fermion parity in 1D. Consider a ring with $N$ equally spaced sites where site $n\in\left\{-\frac{N}{2}+1,\ldots,\frac{N}{2}\right\}$, and assume $N$ is even. The Jordan-Wigner transformation is given as
\begin{equation}\begin{split}\label{eq:JWapp}
&\sigma_{j}^{+}=\frac{1}{2}\left(\sigma_{j}^{x}+i\sigma_{j}^{y}\right)=c_{j}^{\dagger}K^{\dagger}(j),\\
&\sigma_{j}^{-}=\frac{1}{2}\left(\sigma_{j}^{x}-i\sigma_{j}^{y}\right)=K(j)c_{j},\\
&\sigma_{j}^{z}=2n_{j}-1,
\end{split}\end{equation}
where $n_{j}=c_{j}^{\dagger}c_{j}$ is the fermion number operator at site $j$, and $K(j)$ is a string operator defined in the Pauli basis as,
\begin{equation}\begin{split}\label{eq:JWappII}
&K(j)=\exp\left\{i\pi\sum\limits_{\ell=-\frac{N}{2}+1}^{j-1}\sigma_{\ell}^{+}\sigma_{\ell}^{-}\right\}\\
&=(-1)^{j-1}\prod\limits_{\ell=-\frac{N}{2}+1}^{j-1}\sigma_{\ell}^{z},
\end{split}\end{equation}
and in the fermionic basis as,
\begin{equation}\begin{split}\label{eq:JWappIII}
&K(j)=\exp\left\{i\pi\sum\limits_{\ell=-\frac{N}{2}+1}^{j-1}n_{\ell}\right\}\\
&=(-1)^{j-1}\prod\limits_{\ell=-\frac{N}{2}+1}^{j-1}(2n_{\ell}-1).
\end{split}\end{equation}
The total fermion number $\mathcal{N}_{F}$ operator is defined as
\begin{equation}\begin{split}\label{eq:fermionnumberapp}
\mathcal{N}_{F}=\sum\limits_{k=-\frac{N}{2}+1}^{\frac{N}{2}}n_{j}\in\mathbb{Z}_{\geq 0}.
\end{split}\end{equation}
When periodic/anti-periodic boundary conditions are present, the fermion parity becomes a vital consideration, which can be defined as follows,
\begin{equation}\begin{split}\label{eq:fermionparityapp}
\mathcal{P}=\exp\left\{i\pi\mathcal{N}_{F}\right\}=\prod\limits_{k=-\frac{N}{2}+1}^{\frac{N}{2}}\sigma_{k}^{z}.
\end{split}\end{equation}
This implies that when the number of fermions is even ($\mathcal{N}_{F}\equiv 0\hspace{0.1cm}(\text{mod}\hspace{0.1cm}2)$), $\mathcal{P}=+1$, whereas when the number of fermions is odd ($\mathcal{N}_{F}\equiv 1\hspace{0.1cm}(\text{mod}\hspace{0.1cm}2)$), $\mathcal{P}=-1$. Note that $\mathcal{N}_{F}$ is only well-defined with $U(1)$ symmetry is present, as this symmetry enforces charge conservation. Otherwise, if $U(1)$ is broken but $\mathbb{Z}_{2}$ symmetry is present, \eqref{eq:fermionparityapp} still holds but $\mathcal{N}_{F}$ is no longer conserved.

\section{Initial State Protocol for the XX Model}\label{sec:XXmodelprep}

In this appendix, we provide the full details of the algorithm to prepare the particle-hole excited states with fixed $(q,\mathcal{N}_{F})$ for the XX model. This protocol are relevant for the construction of the eigenstate preparation circuit $\mathcal{U}_{\text{PH}}$.\\

\noindent\underline{\textit{Initial State Protocol for XX Model}}

\begin{enumerate}
\setcounter{enumi}{-1}
\item\textit{Momentum Grid Mapping:} The allowed momentum values $k$ depend on whether anti-periodic boundary conditions (APBC) ($\mathcal{N}_{F}\equiv 0\hspace{0.1cm}(\text{mod}\hspace{0.1cm}2)$ or $\mathcal{P}=+1$) or periodic boundary conditions (PBC) ($\mathcal{N}_{F}\equiv 1\hspace{0.1cm}(\text{mod}\hspace{0.1cm}2)$ or $\mathcal{P}=-1$) are present. A typical $N$-qubit computational basis state is expressed as $|q_{0},\ldots,q_{N-1}\rangle$ where $q_{j}\in\{0,1\}$. Before proceeding with the protocol, we first establish a parity-aware mapping $k_{\mathcal{P}}(m):\mathbb{Z}_{\geq 0}\to\mathbb{R}$ between nominal qubit indices and the allowed $k$ values depending on boundary conditions. These mappings are given as 
\begin{equation}\begin{split}
&k_{\mathcal{P}=+1}(m)=\frac{\pi}{N}(2m-(N-1))\hspace{0.25cm}\\
&k_{\mathcal{P}=-1}(m)=\frac{\pi}{N}(2m-(N-2))
\end{split}\end{equation}
where $m\in\left\{0,\ldots,N-1\right\}$. Note that for even $\mathcal{N}_{F}$, the set of allowed momentum values excludes $k=0$ and $k=\pi$, whereas for odd $\mathcal{N}_{F}$, the set of allowed momentum values includes $k=0$ and $k=\pi$.
\item\textit{Qubit Register Initialization:} Initialize all $N$ qubits in the $|0\rangle^{\otimes N}$ state. This represents the fermionic Fock vacuum state.
\item\textit{Filled Fermi Sea Construction for $\mathcal{N}_{F}\in\{2,\ldots,\tfrac{N}{2}\}$:}
This step realizes the Slater determinant state corresponding to the filled Fermi sea state with fixed fermion number $\mathcal{N}_{F}$, such as the ground state ($\mathcal{N}_{F}=N/2$ and $q=0$). With $E(k)=\cos k$ minimized at $k=\pi$, fill the lowest energy modes nearest to $\pi$ on the parity-appropriate grid. Write $\mathcal{N}_F=2r$ (even, APBC) or $\mathcal{N}_F=2r{+}1$ (odd, PBC), $1\leq r\leq\frac{N}{4}$. The occupied momenta are
\begin{equation}\begin{split}
\mathcal{K}_{\text{FS}}=\left\{\pm\left(\pi-(2j+1)\frac{\pi}{N}\right)\right\}_{j=0}^{r-1},
\end{split}\end{equation}
for $\mathcal{N}_{F}=2r$, and 
\begin{equation}\begin{split}
\mathcal{K}_{\text{FS}}=\left\{-\pi+\frac{2\pi}{N}j,\pi+\frac{2\pi}{N}j\right\}_{j=1}^{r}\cup\left\{\pi\right\},
\end{split}\end{equation}
for $\mathcal{N}_{F}=2r+1$. Using the parity aware map $k_{\mathcal{P}}(m)$, this corresponds to applying X gates on the following qubits with indices,
\begin{equation}\begin{split}
\mathcal{M}_{\text{FS}}=\left\{j,N-1-j\right\}_{j=0}^{r-1}
\end{split}\end{equation}
for $\mathcal{N}_{F}=2r$ and
\begin{equation}\begin{split}
\mathcal{M}_{\text{FS}}=\left\{j-1,N-1-j\right\}_{j=1}^{r}\cup\left\{N-1\right\}
\end{split}\end{equation}
for $\mathcal{N}_{F}=2r+1$. For even $\mathcal{N}_{F}$, the total momentum of the filled Fermi sea state is $q=0$, whereas for odd $\mathcal{N}_{F}$, the total momentum is $q=\pi$.
\item\textit{Construct $\left(q,\mathcal{N}_{F}\right)$ (\textit{single} particle-hole (PH) excitation) state:} This step prepares a $\left(q,\mathcal{N}_{F}\right)$ state corresponding to a PH excitation for all $q\neq 0$ when $\mathcal{N}_{F}$ is even and all $q\neq\pi$ when $\mathcal{N}_{F}$ is odd. Starting with the filled Fermi sea state prepared in the previous step with occupied-index set $\mathcal{M}_{\text{FS}}$ and parity-aware momentum map $k_{\mathcal{P}}(m)$, fix the total momentum $q\in\left\{-\frac{\pi}{N}\left(N-2\right),\ldots,\pi\right\}$ (include $q=0$ if $\mathcal{N}_{F}$ is odd or $q=\pi$ if $\mathcal{N}_{F}$ is even) and define the integer shift
\begin{equation}\begin{split}
s\equiv\frac{N}{2\pi}q\hspace{0.1cm}(\text{mod}\hspace{0.1cm}N).
\end{split}\end{equation}
Denote $k_{p}\equiv k_{\mathcal{P}}(m_{p})$ and $k_{h}\equiv k_{\mathcal{P}}(m_{h})$. Choose a ``hole" index $m_{h}\in\mathcal{M}_{\text{FS}}$ (the edge mode near the Fermi momenta $k_{F}=\pm\pi/2$ is the natural choice) and set the ``particle" index to $m_{p}\equiv (m_{h}+s)\hspace{0.1cm}(\text{mod}\hspace{0.1cm}N)$. Maintaining the $X$ gate placement for the filled Fermi sea construction except for $q_{m_{h}}$, apply an $X$ gate to $q_{m_{p}}$ (adds the fermion). This produces a single PH excitation with
\begin{equation}\begin{split}
k_{p}-k_{h}\equiv q\hspace{0.1cm}(\text{mod}\hspace{0.1cm}2\pi)
\end{split}\end{equation}
so the state lies in the total momentum sector $q$ within the fermion number sector $\mathcal{N}_{F}$.
\item\textit{Entangle Degenerate PH States:} PH symmetry maps $c_{k}^{\dagger}\to c_{\pi-k}$ for the lattice-regularized XX model, resulting in degenerate states for all single PH states with $q\neq 0,\frac{2\pi}{N}$ in the even $\mathcal{N}_{F}$ sector. In the odd $\mathcal{N}_{F}$ sector, only the $q=0$ PH state is degenerate under PH symmetry. For a given PH excitation arising from the Fermi sea created via a hole at $k_{h}$ and an occupied fermion at $k_{p}$, there exists another state equal in energy to a PH excitation with a hole at $\pi-k_{h}$ and fermion occupation at $\pi-k_{p}$. The PH symmetry can be preserved by initializing the state as an equal symmetric/anti-symmetric superposition of these degenerate states, i.e. a Bell/EPR state. Define the PH-partner indices $m_{h}'$ and $m_{p}'$ by
\begin{equation}\begin{split}
&k_{\mathcal{P}}(m_{h}')=\pi-k_{h}\hspace{0.25cm}\&\\
&k_{\mathcal{P}}(m_{p}')=\pi-k_{p}\hspace{0.1cm}(\text{mod}\hspace{0.1cm}2\pi).
\end{split}\end{equation}
Thus, the pair of qubits $q_{m_{p}}$ and $q_{m_{h}}$ correspond to one PH excitation, and the PH-symmetric partner excitation is given by the pair of qubits $q_{m_{p}'}$ and $q_{m_{h}'}$. One can curate a symmetric/anti-symmetric Bell/EPR pair state between these degenerate states through the following procedure:
\begin{itemize}
\item Initialize Fermi Sea state from Step 2, but without an $X$ gate applied to the qubit $q_{\ell_{h}}$ where $\ell_{h}=\min(m_{h},m_{h}')$
\item Create PH state for the pair of qubits $q_{\ell_{h}}$ \& $q_{\ell_{p}}$ following Step 3, where $\ell_{p}$ is the corresponding particle index (either $m_{p}$ or $m_{p}'$) depending on the value of $\ell_{h}$.
\item Apply a Hadamard (H) gate on the qubit $q_{\ell_{h}}$.
\item Apply a controlled-NOT (CNOT) gate to the qubits $q_{\ell_{h}}$ (control) and $q_{\bar{\ell}_{h}}$ (target), where $\bar{\ell}_{h}=\max(m_{h},m_{h}')$.
\item Apply a CNOT gate to the qubits $q_{\ell_{h}}$ (control) and $q_{m_{p}}$ (target).
\item Apply a CNOT gate to the qubits $q_{\ell_{h}}$ (control) and $q_{m_{p}'}$ (target).
\item (Optional) Apply a Z gate to $q_{\ell_{h}}$ to create an anti-symmetric superposition.
\end{itemize}
An example of the circuit construction for a superposition of degenerate PH pair of states is shown in Fig.~\ref{fig:PHCircuitapp}, for an $N=8$ qubit system. This is a $q=\frac{\pi}{2}$ state, constructed by taking a superposition between two PH states: one with $(k_{p},k_{h})=\left(-\frac{\pi}{8},-\frac{5\pi}{8}\right)$ and another with $(k_{p}',k_{h}')=\left(-\frac{7\pi}{8},-\frac{3\pi}{8}\right)$. First, we identify the hole indices, which are $m_{h}=0$ and $m_{h}'=1$, which means $\ell_{h}=0$. An $H$ gate is then applied $q_{0}$. Next, we identify the particle indices, which are $m_{p}=2$ and $m_{p}'=3$. A CNOT is first applied to the qubits $q_{0}$ (control) and $q_{1}$ (target), followed by another CNOT applied to qubits $q_{0}$ (control) and $q_{2}$ (target), followed by one last CNOT applied to qubits $q_{0}$ (control) and $q_{3}$ (target).
\end{enumerate}

\begin{figure}[htbp]
\includegraphics[width=0.35\textwidth]{"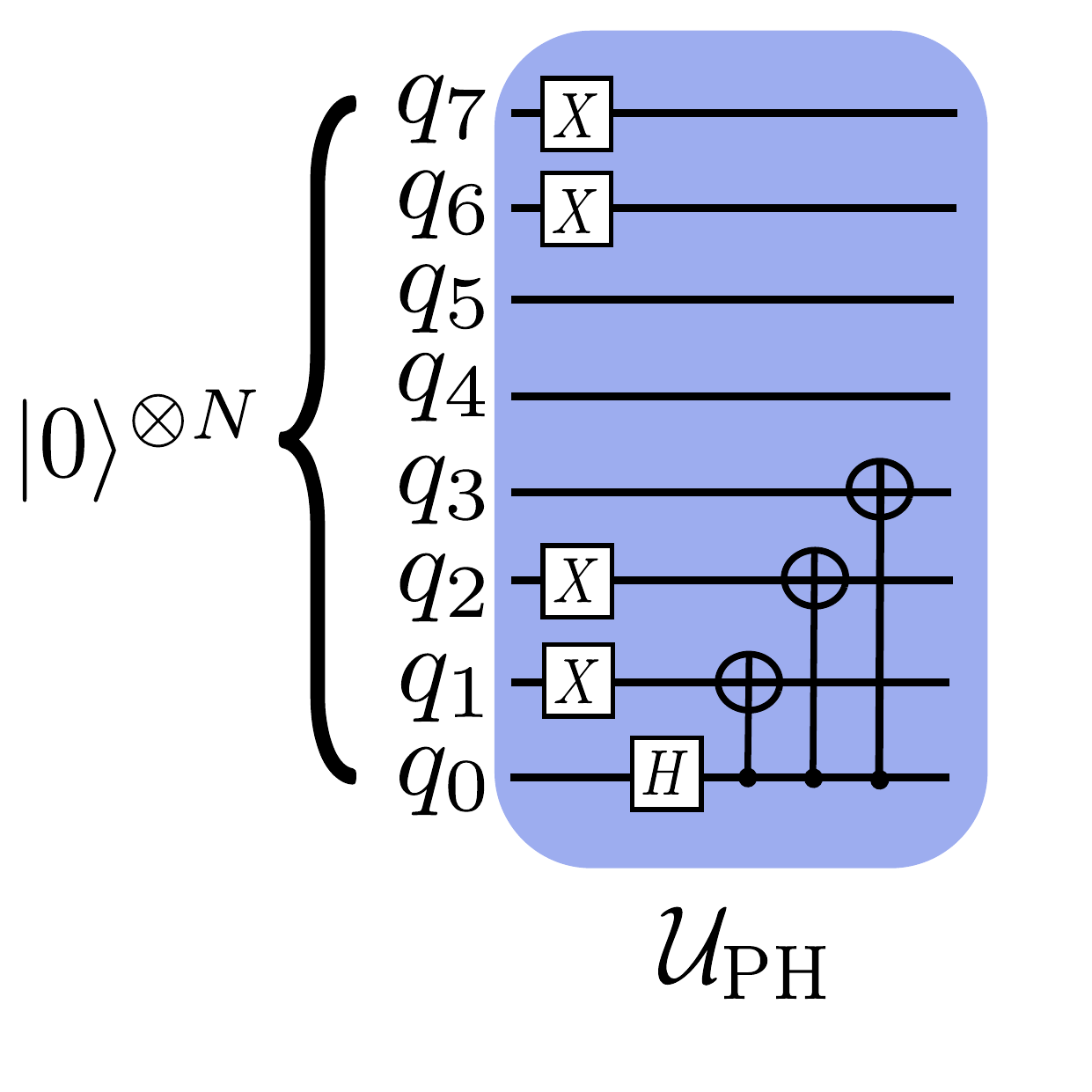"}
\caption{Eigenstate preparation circuit $\mathcal{U}_{\text{PH}}$ for the construction of an Bell/EPR pair state corresponding to a superposition of PH states with total momentum $q=\frac{\pi}{2}$ in the $\mathcal{N}_{F}=4$ sector for an $N=8$ qubit system.}
\label{fig:PHCircuitapp}
\end{figure} 

\section{Alternate Symmetry-Preserving HVA Construction}\label{sec:alternateHVA}

In this appendix, we review an alternate symmetry-preserving HVA construction proposed in Ref.~\cite{Babbush2018} that exactly preserves both translation and $U(1)$ symmetries. Note that under the Fourier transform, the nearest-neighbor hopping term in Eq.~\eqref{eq:fermionicHam} is diagonalized. Using Eq.~\eqref{eq:FFFTunitary}, one can express this term as
\begin{equation}\begin{split}
2\sum\limits_{n}\left(c_{n}^{\dagger}c_{n+1}+\text{h.c.}\right)=\mathcal{U}_{\text{FFFT}}^{\dagger}\left(\sum\limits_{k}\epsilon(k)\tilde{c}_{k}^{\dagger}\tilde{c}_{k}\right)\mathcal{U}_{\text{FFFT}}.
\end{split}\end{equation}
This means one can rewrite Eq.~\eqref{eq:fermionicHam} in the main text as follows,
\begin{equation}\begin{split}\label{eq:fermionicHamII}
&\mathcal{H}_{\text{XXZ}}=\mathcal{U}_{\text{FFFT}}^{\dagger}\left(\sum\limits_{k}\epsilon(k)\tilde{c}_{k}^{\dagger}\tilde{c}_{k}\right)\mathcal{U}_{\text{FFFT}}-\mu\sum\limits_{n}c_{n}^{\dagger}c_{n}\\
&+U\sum\limits_{n}c_{n}^{\dagger}c_{n+1}^{\dagger}c_{n+1}c_{n}.
\end{split}\end{equation}
In this representation, the nearest-neighbor hopping term is diagonalized in momentum space ($k$ is a momentum index, where the set of allowed momentum values is determined by fermion parity), while the chemical potential and interaction terms are still expressed in terms of fermionic number operators in real space ($n$ is the site index). Employing the Jordan-Wigner transformation, Eq.~\eqref{eq:fermionicHamII} becomes (neglecting overall constants)
\begin{equation}\begin{split}\label{eq:fermionicHamIII}
&\mathcal{H}_{\text{XXZ}}=\mathcal{U}_{\text{FFFT}}^{\dagger}\left(\frac{1}{2}\sum\limits_{n}\epsilon\left(k_{\mathcal{P}}(n)\right)\sigma_{n}^{z}\right)\mathcal{U}_{\text{FFFT}}\\
&-\left(\frac{\mu}{2}-\frac{U}{2}\right)\sum\limits_{n}\sigma_{n}^{z}+\frac{U}{4}\sum\limits_{n}\sigma_{n}^{z}\sigma_{n+1}^{z}.
\end{split}\end{equation}
where we have used the parity-aware fermion map given by Eq.~\eqref{eq:fermionparityaware} and expressed the sum in the first term over the qubit indices $n\in\left\{0,\ldots,N-1\right\}$. Using this observation, one can construct the HVA based on the three terms present in Eq.~\eqref{eq:fermionicHamIII}, which yields Eq.~\eqref{eq:HVAIII}. Note that since $\mu=U=4\Delta$, the Hartree shift term is cancelled out, which can reduce Eq.~\eqref{eq:HVAIII} to a $(N+1)p$-parameter ansatz.

\section{Eigenstate Indices for ED Results}\label{sec:tables}

In this appendix, we provide the indices of the lowest energy eigenstates in the spectrum of $\mathcal{H}_{\text{XXZ}}$ labeled by the quantum numbers $\left(q,\mathcal{N}_{F}\right)$ where $q$ is the total momentum and $\mathcal{N}_{F}$ is the fermion number.

\begin{widetext}
\noindent
\begin{minipage}[t]{0.48\textwidth}
\centering
\textbf{$N=8$ Qubits}\\[2pt]
\renewcommand{\arraystretch}{1}
\begin{tabular}{cc} 
\hline
$(q,N_{F})$ & Many-Body Eigenstate Index (Indices) \\
\hline
$(0,4)$         & 0 \\
$(\pi,3)$       & 1, 2 \\
$(\pi,4)$       & 3, 4 \\
$(\pm\pi/4,4)$     & 5, 6 \\
$(\pm\pi/4,3)$     & 7, 8, 9, 10 \\
$(\pm 3\pi/4,3)$    & 11, 12, 13, 14 \\
$(\pm 3\pi/4,4)$    & 17, 18, 25, 26 \\
$(\pm \pi/2,3)$     & 19, 20, 21, 22 \\
$(\pm \pi/2,4)$     & 23, 24, 27, 28 \\
$(0,3)$         & 29, 30, 47, 48 \\
\hline
\end{tabular}
\end{minipage}\hfill
\begin{minipage}[t]{0.48\textwidth}
\centering
\textbf{$N=16$ Qubits}\\[2pt]
\renewcommand{\arraystretch}{1.05}
\begin{tabular}{cc} 
\hline
$(q,N_{F})$ & Many-Body Eigenstate Index (Indices) \\
\hline
$(0,8)$        & 0 \\
$(\pi,7)$      & 1, 2 \\
$(\pi,8)$      & 3, 4 \\
$(\pm \pi/8,8)$    & 5, 6 \\
$(\pm \pi/8,7)$    & 7, 8, 9, 10 \\
$(\pm 7\pi/8,7)$   & 11, 12, 13, 14 \\
$(\pm 7\pi/8,8)$   & 17, 18, 19, 20 \\
$(\pm \pi/4,8)$    & 21, 22, 25, 26 \\
$(0,7)$        & 23, 24, 40, 41 \\
$(\pm \pi/4,7)$    & 27, 28, 29, 30 \\
$(\pm 3\pi/4,7)$   & 36, 37, 38, 39 \\
$(\pm 3\pi/4,8)$   & 52, 53, 55, 56 \\
$(\pm 3\pi/8,7)$   & 57, 58, 59, 60 \\
$(\pm 3\pi/8,8)$   & 61, 62, 65, 66 \\
$(\pm 5\pi/8,7)$   & 67, 68, 69, 70 \\
$(\pm \pi/2,7)$    & 82, 83, 84, 85 \\
$(\pm 5\pi/8,8)$   & 86, 87, 108, 109 \\
$(\pm \pi/2,8)$    & 92, 93, 117, 118 \\
\hline
\end{tabular}
\end{minipage}
\end{widetext}
      
\end{document}